\documentclass[12pt,english,preprint]{revtex4}
\usepackage[utf8]{inputenc}
\usepackage{array}
\usepackage{longtable}
\usepackage{float}
\usepackage{amsmath}
\usepackage{graphicx}
\usepackage{amssymb}
\usepackage{color}
\usepackage[FIGTOPCAP]{subfigure}

\begin{document}

\title{Physics-Constrained Deep Learning of Incompressible Cavity Flows}

\author{Christopher J. McDevitt}
\address{University of Florida}
\author{Eric Fowler}
\address{University of Florida}
\author{Subrata Roy}
\address{University of Florida}

\date{\today}

\begin{abstract}

High resolution simulations of incompressible Navier-Stokes flows have become routine across a range of engineering applications. Despite their routine use, due to the high dimensional parameter space present for most practical applications, a comprehensive exploration of the available parameter space is often impractical. In this work, we demonstrate the ability of physics-constrained deep learning methods to provide an efficient means of exploring high-dimensional parameter spaces with minimal amounts of data from high resolution computational fluid dynamic simulations. As a specific application, we choose the well established problem of a two-dimensional lid driven cavity flow. While giving an extensive treatment of the classic case of a square cavity, we extend the analysis to treat an isosceles trapezoid. In so doing, the number of parameters determining the solution includes not just the Reynolds number, but also two additional parameters characterizing the geometry of the cavity. Thus, together with the $\left( x, y \right)$ variation of the flow and pressure in configuration space, the presence of these three parameters results in the solution varying in a five-dimensional space. 
It is shown that in the absence of data, physics-constrained methods are able to provide an accurate description of the cavity flow in this five-dimensional space up to intermediate values of the Reynolds number, but fails to train for sufficiently high Reynolds numbers. In contrast, using a small quantity of flow data, a single neural network is able to provide an accurate description for a broad range of Reynolds numbers and cavity geometries. Once trained, such a model provides a rapid surrogate model for predicting the flow structure and can thus be used to efficiently explore the five dimensional space. This five-dimensional surrogate model is subsequently used to identify critical parameter values for the merger and splitting of vortices as the Reynolds number and cavity geometry are varied.

\end{abstract}

\maketitle

\section{Introduction}

Computational fluid dynamic (CFD) simulations have become an indispensable tool across a range of engineering and scientific applications. Despite their central role, the employment of these tools is often complicated by the need to incorporate complex geometries, multiphysics applications, and the treatment of real time or many-query analyses. Due to these limitations, several authors have pursued the use of artificial neural networks (NN) to overcome several of the most challenging aspects of CFD simulations. Originally proposed in 1943 \cite{mcculloch1943logical}, artificial NNs has since become wildly popular for a wide variety of applications with sophisticated algorithms and increasingly faster graphical and standard processing units. Machine learning (ML) applications to fluid dynamics include the development of Reynolds averaged Navier-Stokes (RANS) turbulence models~\cite{ling2016reynolds}, flow control~\cite{mohan2018deep, ren2021applying}, and model extraction~\cite{brunton2016discovering}.

Recent innovations in physics-informed deep learning methods have suggested the tantalizing possibly of overcoming several of the obstacles traditionally faced by CFD simulations~\cite{lagaris1998artificial,karpatne2017theory, karniadakis2021physics, lusch2018deep, wang2020towards}. Broadly speaking, physics informed deep learning approaches seek to incorporate physical constraints into the training of an artificial neural network (NN).
This can be done by either modifying the NN architecture to only output solutions consistent with a given physical constraint, a method referred to as inductive bias, or via the introduction of a penalty function that pushes the solution toward the desired physical constraint.
While the former approach offers the benefit of encoding hard constraints into the NN, and thus drastically reduces the space of available solutions, the latter approach often offers greater flexibility in terms of the types of physical constraints that can be enforced. A particularly attractive example of the latter approach corresponds to Physics-Informed Neural Networks (PINNs)~\cite{raissi2019physics}, whereby physical constraints are embedded in the loss function of the NN, in the form of symmetries, conservation laws, or explicit PDEs. This approach, along with its relatives~\cite{wang2021learning}, offers the possibility of alleviating several of the challenges associated with CFD simulations. In particular, this approach does not require a numerical mesh, and can thus straightforwardly treat complex geometries without the laborious task of mesh generation. Multiphysics problems can be treated by combining surrogate models for each physics component, yielding an efficient multiphysics framework~\cite{mao2021deepm}. Furthermore, PINNs have been shown to be effective at learning the parameter space of a PDE, thus providing a rapid surrogate model capable of treating many-query analyses, a central component to the treatment of design, optimization, and uncertainty quantification problems~\cite{sun2020surrogate}. The primary limitation of this approach, however, is that PINNs to date have not been able to compete with CFD solvers with regard to obtaining high resolution solutions of high Reynolds number flows~\cite{karniadakis2021physics} in the absence of synthetic or experimental data, thus sharply limiting their application.

In the present work we will explore a paradigm through which PINNs are employed as a complementary tool to CFD solvers for the purpose of efficiently exploring high dimensional parameter spaces. Namely, by exploiting the ease in which data can be integrated into the training of PINNs we will seek to train a model capable of describing high Reynolds number flows that would otherwise require high resolution CFD solvers. Aside from simply learning this CFD data, however, the physical constraints embedded in the training of the NN will enable the model to make predictions across a range of Reynolds numbers and geometries where CFD data is not available. In so doing we aim to retain the most attractive aspects of high resolution CFD solvers, specifically their ability to accurately describe high Reynolds number flows, while utilizing PINNs to track how features of the solution change as the parameters or problem geometry is varied.

As a specific realization of this approach, we will develop a surrogate model capable of learning the flow and pressure variation for the classic problem of a viscous flow in a lid driven cavity \cite{ghia1982high}. The flow physics in a lid driven cavity, albeit incompressible, provide the characteristic singularities from the velocity discontinuity at geometric corners resulting in flow instabilities which can eventually transition to chaos in a fully confined cubical cavity ~\cite{kuhlmann2018}. Smart predictive capability for this problem will set the stage for theoretical understanding of canonical cases for studying various instabilities including the elliptic instability, and the quadripolar instability (due to a Kelvin-wave resonance) communicated by a quadripolar strain field \cite{eloy2001}. It has been shown that for the resonance condition the critical wave number in the lid-driven cavity flow is quite large ($k_c\approx15$) ~\cite{kuhlmann2018}. For reference, Chandrasekhar described that the columnar vortex \cite{chandra1961} can only be satisfied for asymptotically large wave number $k$. 

Particular attention will be given to the classic problem of a square cavity. However, to demonstrate the ability of the present approach to treat high dimensional parameter spaces, and nontrivial geometries, we will also treat the case of a trapezoidal cavity (see Fig. \ref{fig:Cavity}). By treating a trapezoidal cavity this introduces two parameters characterizing the geometry of the cavity. Thus, aside from the two dimensions in configuration space, $\left( x,y\right)$, the problem is parameterized by the Reynolds number $Re$, the aspect ratio of the cavity $K\equiv D/L$, where $D$ and $L$ are the depth and width of the cavity respectively, and the parameter $a\equiv \delta / L$ that determines the difference in length between upper and lower lengths of the trapezoid (see Fig. \ref{fig:Cavity}). Thus, once trained the NN will be able to provide rapid predictions of the flow and pressure as a function of the five-dimensional space spanned by $\left(x,y,Re,K, a\right)$. 

The remainder of this paper is organized as follows. Section \ref{sec:PCDL} describes the PINN formulation employed. Subsequently, Sec. \ref{sec:LDCF} defines two different formulations of the steady state incompressible Navier-Stokes equation, and compares the performance of PINNs when describing the flow inside a square cavity. The specific case of developing a surrogate model for a square cavity at high Reynolds numbers is treated in Sec. \ref{sec:SC}, whereas the trapezoidal cavity is described in Sec. \ref{sec:TC}. Conclusions and a discussion of the observed strengths and weaknesses of the framework are provided in Sec. \ref{sec:C}.

\section{Methodology}


\subsection{\label{sec:PCDL}Physics-Constrained Deep Learning}

The use of physical constraints for training a neural network has recently received a substantial amount of attention across a range of applications, with several excellent reviews available (see Ref. \cite{karniadakis2021physics}, for example). Here we briefly describe some critical aspects of this approach that will be exploited in the present work. The PINN framework is focused on minimizing the residual of one of more PDEs, along with boundary and initial conditions to obtain a NN that approximates the solution of the targeted PDE(s). A loss function that accomplishes this task can be expressed as~\cite{raissi2019physics}
\begin{equation}
\text{Loss} = \frac{1}{N_{PDE}} \sum^{N_{PDE}}_i  \left| R \left( x_i \right) \right|^2 +  \frac{\lambda}{N_{bdy}} \sum^{N_{bdy}}_i \left| u \left( x_i \right) - u_i \right|^2
, \label{eq:PCDL0}
\end{equation}
where $R \left( x \right)$ represents the residual of the PDE(s), the boundary condition on the solution $u$ is enforced by the second term, $\lambda$ is a parameter that may be adjusted to determine the relative importance of the PDE and boundary penalty terms. A steady state problem was assumed for simplicity, and $N_{PDE}$ and $N_{bdy}$ represent the number of points to sample inside the simulation domain and along the boundary, respectively. For a time dependent problem, an additional penalty term would need to be included to enforce the initial condition. An approximate solution to the PDE is obtained by solving an optimization problem whereby the weights and biases of the NN that minimize Eq. (\ref{eq:PCDL0}) are identified. This approach, while capable of solving several PDEs of interest to scientific and engineering applications, suffers from several limitations. The first is that for a multidimensional problem, with multiple fields, this leads to a large number of terms in the loss function, and hence a  multi-objective optimization problem that may fail to train all together, or have difficulty identifying the global minimum~\cite{wang2022and}. In practice tuning the relative weighting of the terms of the loss function (i.e. $\lambda$) is often required, either manually, or via an adaptive scheme~\cite{jin2021nsfnets}. This limitation can be largely overcome in several situations of interest by enforcing boundary conditions as hard constraints~\cite{lagaris1998artificial, sheng2021pfnn}, thus removing the need to include the second term in Eq. (\ref{eq:PCDL0}). For the case of Dirichelet boundary conditions, this can be done by modifying the output of the NN by~\cite{lu2021physics}:
\begin{equation}
u \left( x \right) = u_{bdy} \left( x \right) + u_{zero} \left( x\right) u^\prime \left( x \right)
, \label{eq:PCDL0sub1} 
\end{equation}
where $u^\prime \left( x \right)$ is the output of the neural network and $u \left( x \right)$ is the predicted solution (flow velocity, for example). Here, the function $u_{zero} \left( x\right)$ is defined such that it vanishes on the system's boundaries and $u_{bdy} \left( x \right)$ is defined to satisfy the boundary conditions. In this way, regardless of the output of the NN (i.e. the value of $u^\prime$) the boundary conditions are automatically satisfied. This approach removes the need for the second term in Eq. (\ref{eq:PCDL0}), thus reducing the number of terms in the loss function to simply the residual of the PDE(s).

An additional challenge faced by PINNs is that they often struggle to identify complex flow solutions present at high Reynolds numbers. This is discussed in Sec. \ref{sec:SC} below. This limitation can be relaxed by the inclusion of a modest quantity of CFD or experimental data. To incorporate data into the training of the NN, an additional term is added to the loss function yielding
\begin{equation}
\text{Loss} = \frac{1}{N_{PDE}} \sum^{N_{PDE}}_i  \left| R \left( x_i \right) \right|^2 +  \frac{\lambda}{N_{data}} \sum^{N_{data}}_i \left| u \left( x_i \right) - u_i \right|^2
. \label{eq:PCDL0sub2}
\end{equation}
Here, $N_{data}$ represents the number of data points available, we have dropped the boundary penalty term since it is enforced as a hard constraint, and added a term expressing the loss with respect to available CFD or experimental data. As demonstrated in Sec. \ref{sec:SC}, the inclusion of a modest amount of flow data at specific points in the cavity greatly extends the ability of the NN to accurately predict the dominant global flow structures. 

The final feature of this approach that we will exploit will be the relative ease that NNs are able to learn high dimensional solutions~\cite{sun2020surrogate}. Thus, rather than learning the solution for a given Reynolds number and cavity geometry we will seek to learn the flow and pressure profiles for a broad range of Reynolds numbers and cavity geometries. The loss function will thus be modified to take the form
\begin{equation}
\text{Loss} = \frac{1}{N_{PDE}} \sum^{N_{PDE}}_i  \left| R \left( x_i,\mathbf{\theta}_i \right) \right|^2 +  \frac{\lambda}{N_{data}} \sum^{N_{data}}_i \left| u \left( x_i,\mathbf{\theta}_i \right) - u_i \right|^2
, \label{eq:PCDL0sub3}
\end{equation}
where $\mathbf{\theta}$ are the parameters of the PDE (Reynolds number) along with additional parameters related to the cavity geometry ($K, a$). It will be shown that the inclusion of small quantities of data at specific Reynolds numbers and cavity geometries aid in the training of the NN across a broad range of Reynolds numbers and cavity geometries. 

For all studies carried out in this work we choose a fully connected feed-forward deep neural network, initialized with a Glorot normal distribution. The training points $N_{PDE}$ are randomly drawn from a uniform distribution across the flow cavity where a sufficient number of training points are used to make the results reproducible. CFD flow data will only be utilized when explicitly mentioned. A tanh activation function is employed throughout. Finally we note that several outstanding libraries are available for employing physics informed approaches to deep learning including DeepXDE~\cite{lu2021deepxde} and SimNet (more recently renamed Modulus)~\cite{hennigh2021nvidia}. For the present study we have elected to use DeepXDE. The python codes used for the primary studies carried out in this work will be made available through GitHub after acceptance of this article.

\subsection{\label{sec:LDCF}Formulations of Incompressible Navier-Stokes Equations}

\begin{figure}
\begin{centering}
\includegraphics[scale=0.45]{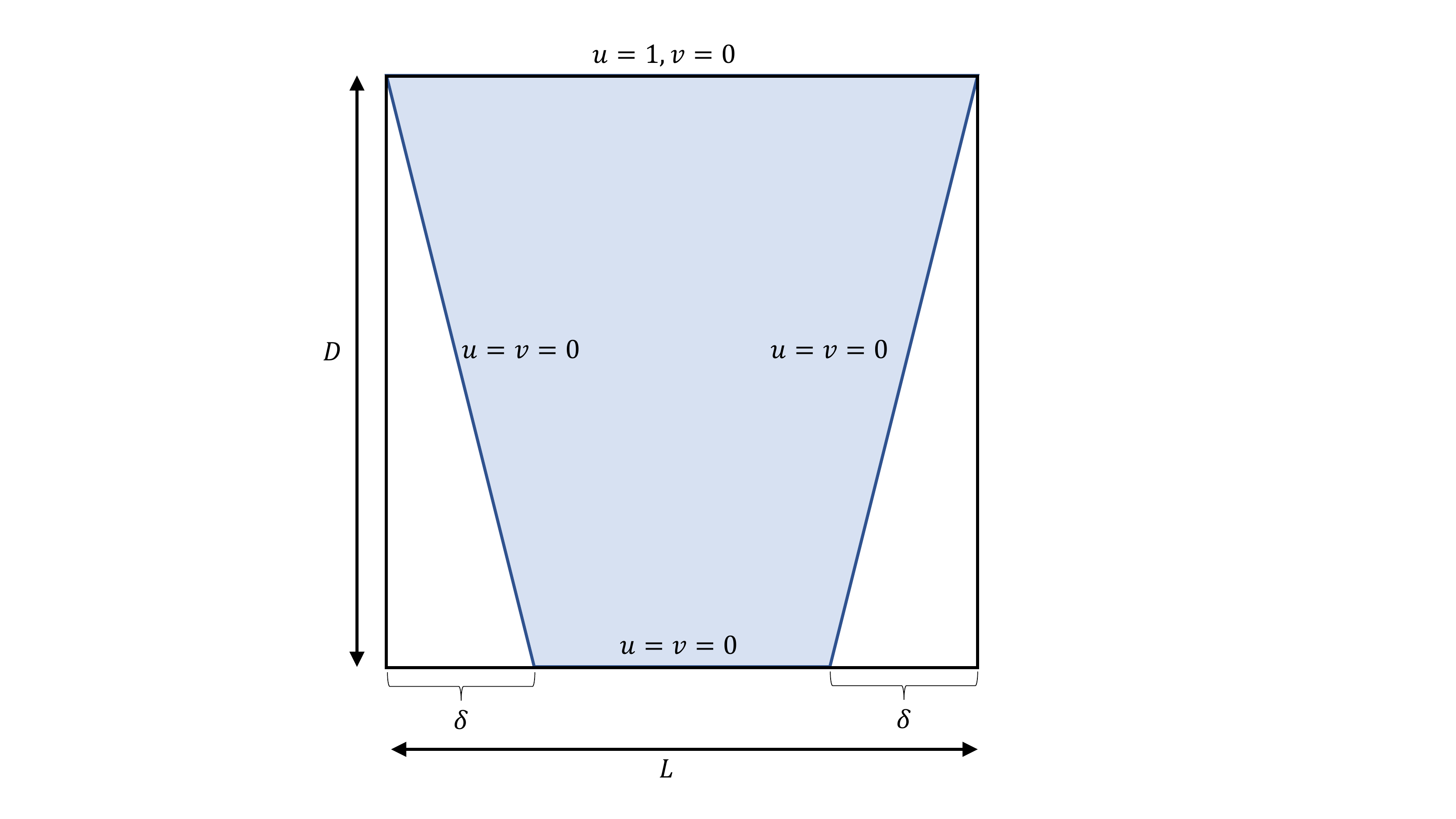}
\par\end{centering}
\caption{Cavity geometry. The blue shaded region represents the simulation domain.}
\label{fig:Cavity}
\end{figure}

The incompressible Navier-Stokes equation may be expressed by:
\begin{subequations}
\label{eq:2DLD1} 
\begin{equation}
\mathbf{u} \cdot \nabla \mathbf{u} = -\nabla p +\frac{1}{Re} \nabla^2 \mathbf{u}
, \label{eq:2DLD1a} 
\end{equation}
\begin{equation}
\nabla \cdot \mathbf{u} = 0
. \label{eq:2DLD1b} 
\end{equation}
\end{subequations}
Here, $Re \equiv u_{lid} L/\nu$ is the Reynolds number, $\nu$ is the kinematic viscosity, $L$ is the width of the cavity, $u_{lid}$ is the speed at which the lid (the upper boundary) of the cavity is sliding, spatial scales have been normalized to $L$, and velocities to $u_{lid}$. The geometry of the cavity is illustrated in Fig. \ref{fig:Cavity}.
When training the network we will be interested in enforcing as many properties of the flow as possible as hard constraints. In so doing, this not only ensures fundamental aspects of the problem are exactly satisfied, but it also reduces the number of terms in the loss function, which facilitates the training of the neural network. In this work we will explore two formulations of the Navier-Stokes equation. In the first approach, which we will refer to as the $\left( v,p\right)$ formulation, we will solve Eq. (\ref{eq:2DLD1}) for $\left( u,v,p\right)$, where the boundary conditions will be enforced as hard constraints. This latter property is achieved by introducing an additional layer to the network, whereby the solution is transformed by~\cite{lu2021physics} 
\begin{subequations}
\label{eq:2DLD2} 
\begin{equation}
u \left( x,y \right) = u_{lid} \left( x,y \right) + u_{zero} \left( x,y\right) u^\prime \left( x,y \right)
, \label{eq:2DLD2a} 
\end{equation}
\begin{equation}
v \left( x,y \right) = v_{zero} \left( x,y\right) v^\prime \left( x,y \right)
, \label{eq:2DLD2b}
\end{equation}
\begin{equation}
p \left( x,y \right) = p^\prime \left( x,y \right)
, \label{eq:2DLD2c} 
\end{equation}
\end{subequations}
where the functions with primes indicate the outputs of the NN, and the function $u_{lid}\left( x,y \right)$, $u_{zero}\left( x,y \right)$, and $v_{zero}\left( x,y \right)$ are chosen to satisfy the flow boundary conditions. For the special case of a square domain we will choose:
\[
u_{zero} \left( x,y\right) = v_{zero} \left( x,y \right) = 16x\left( x-1\right)y\left( y-1\right)
, 
\]
\[
u_{lid} = \sin \left[ \frac{3\pi}{2} \left( y - \frac{2}{3} \right) \right] \left[ 1 - \exp \left(- \frac{\left( x-1 \right)^2}{\Delta x^2} \right) \right] \left[ 1 - \exp \left( -\frac{x^2}{\Delta x^2} \right) \right]
,
\]
where $\Delta x \ll 1$. With these definitions, the flow boundary conditions can be verified to satisfy the boundary conditions throughout the simulation domain as $\Delta x \to 0$. Here, the flow $u_{lid} \left( x,y\right)$ allows the $u=1$ boundary condition at the upper boundary to be satisfied. By introducing the boundary conditions as hard constraints, this will allow us to tune the accuracy of the solver to obtain the best convergence possible for the interior points, without the need to also minimize boundary values (since they are automatically satisfied). The choice of a sine function for $u_{lid}$, was motivated to select a structure for the $u$-component of the flow that is qualitatively similar to the vortical flow that we anticipate to develop. A final note is that the above functions only satisfy the boundary conditions exactly for $\Delta x \to 0$. For finite $\Delta x$, there will be a slight deviation from the boundary conditions near the corners of the simulation domain. For the studies described in this work we will choose $\Delta x^2 = 10^{-3}$.

The second formulation we will employ, which we will refer to as the $\left( \psi, p\right)$ formulation, will enforce incompressibility along with the boundary conditions as hard constraints. To do this, we begin by noting that if the flow is expressed in terms of a stream function $\psi$ defined by $u \equiv \frac{\partial \psi}{\partial y},\quad v \equiv - \frac{\partial \psi}{\partial x}$ incompressibility will be exactly satisfied. Such an approach is commonly used for the $\left( \psi, \omega \right)$ formulation of Navier-Stokes, whereby a vorticity equation together with a Poisson equation for the stream function are solved simultaneously. 
Here, we will seek an intermediate approach. We will still enforce the Navier-Stokes equation in the form Eq. (\ref{eq:2DLD1a}), but an additional layer of the network will be added that will enforce incompressibility. A similar approach was recently employed in Ref. \cite{amalinadhi2022physics}. Explicitly, the output of the NN will be taken to be $\left( \psi^\prime, p^\prime \right)$.
The first outer layer is designed to enforce the boundary conditions of the problem as a hard constraint. This is done by defining the stream function $\psi$ as
\begin{equation}
\psi \left( x, y\right) = \psi_{lid} \left( x, y\right) + \psi_{zero} \left( x, y\right) \psi^\prime \left( x, y\right)
, \label{eq:2DLD2sub2}
\end{equation}
\begin{equation}
p \left( x, y \right) = p^\prime \left( x, y\right)
, \label{eq:2DLD2sub2sub1}
\end{equation}
where $\psi_{lid}$ and $\psi_{zero}$ are defined by:
\begin{equation}
\psi_{lid} \left( x, y\right) \equiv \left( y-1\right)y^2  \left[ 1 - \exp \left(- \frac{\left( x-1 \right)^2}{\Delta x^2} \right) \right] \left[ 1 - \exp \left( -\frac{x^2}{\Delta x^2} \right) \right]
, \label{eq:2DLD2sub3}
\end{equation}
\begin{equation}
\psi_{zero} \left( x, y\right) \equiv 256 x^2 \left( x-1\right)^2 y^2 \left( y-1\right)^2
. \label{eq:2DLD2sub4}
\end{equation}
It can be verified that a stream function $\psi$ defined in this way will satisfy the boundary conditions of vanishing at the simulation boundaries, having a normal derivative of one at the top boundary, and a normal derivative of zero at the other three boundaries.

In order to enforce incompressibility, but still use Eq. (\ref{eq:2DLD1a}) as a constraint, we will introduce a second outer layer which performs the operation 
\begin{equation}
u = \frac{\partial \psi}{\partial y},\quad v = - \frac{\partial \psi}{\partial x}
. \label{eq:2DLD2sub1}
\end{equation}
The output of this layer is then used to evaluate Eq. (\ref{eq:2DLD1a}). While the present approach bears some similarity to the $\left( \psi, \omega \right)$ formulation of Navier-Stokes, significant differences are evident. First, we do need to introduce vorticity. In so doing, we reduce the order of the system of equations that are minimized.
A second distinction is that since $\left( u, v, p\right)$ are the output of the system, flow data can be used to directly train the network, which are often more straightforward to extract from experiments compared to vorticity or a stream function. An additional distinction is that once trained, we will be able to immediately identify the quantities $\left( u, v, p\right)$ along with the stream function $\psi$. This last quantity can be easily extracted in the present formulation since it is generated as an intermediate result when computing $u$ and $v$. This thus provides a practical advantage since there is no need to solve for $\psi$ as part of the post processing of the data. Since incompressibility is automatically satisfied within this second approach we will only enforce Eq. (\ref{eq:2DLD1a}), which thus reduces the number of PDEs appearing in the loss function to two from an initial three.

\begin{figure}
\begin{centering}
\subfigure[]{\includegraphics[scale=0.5]{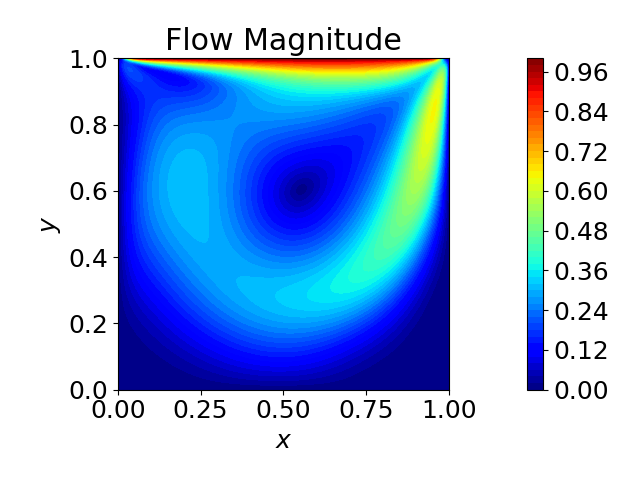}}
\subfigure[]{\includegraphics[scale=0.5]{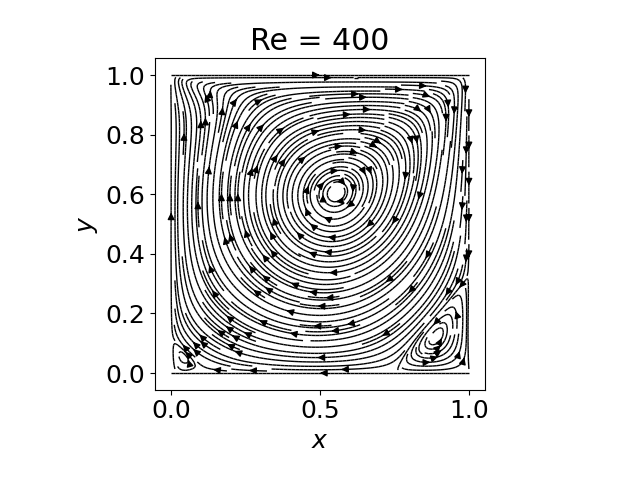}}
\subfigure[]{\includegraphics[scale=0.5]{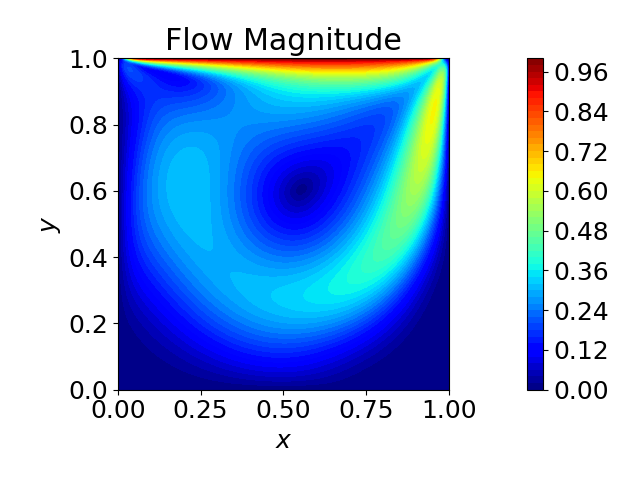}}
\subfigure[]{\includegraphics[scale=0.5]{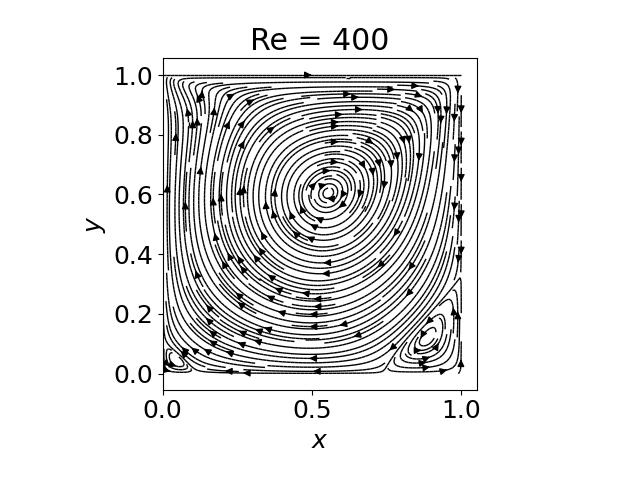}}
\par\end{centering}
\caption{Flow solutions using the $\left( v,p\right)$ (top row) and $\left( \psi, p\right)$ (bottom row) formulations. Flow magnitudes are indicated in the left column and streamlines in the right column. The Reynolds number was taken to be $Re=1000$ and the cavity was taken to have an aspect ratio of $K=1$. A NN with an architecture of $(3) + (32) \times 4 + (3)$ was used, 50,000 training points were employed, with 15,000 iteration with ADAM and 200,000 iterations with L-BFGS.}
\label{fig:LPDSD0}
\end{figure}

\begin{figure}
\begin{centering}
\subfigure[]{\includegraphics[scale=0.33]{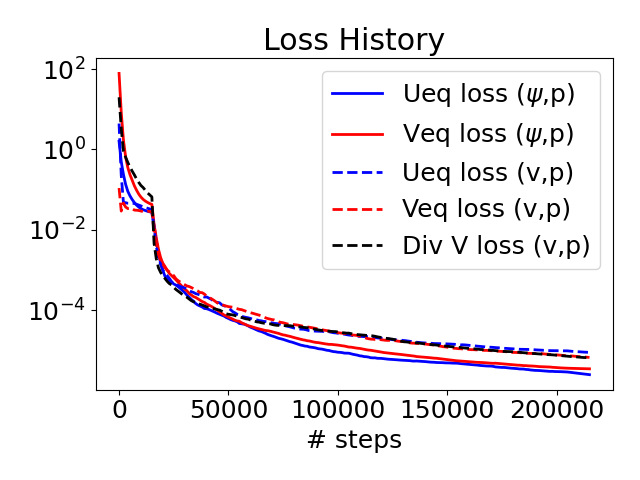}}
\subfigure[]{\includegraphics[scale=0.33]{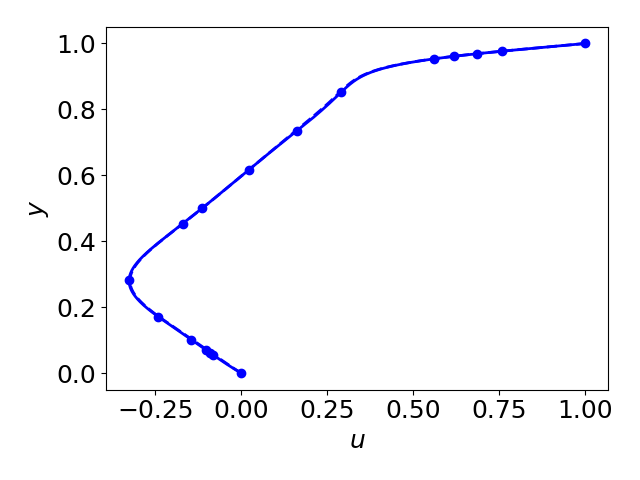}}
\subfigure[]{\includegraphics[scale=0.33]{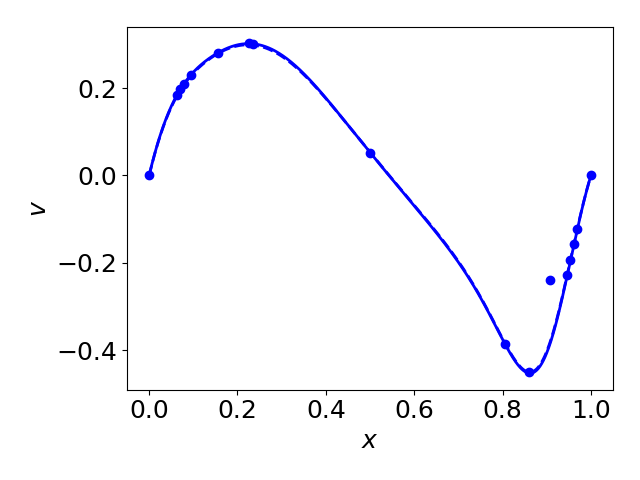}}
\par\end{centering}
\caption{Training loss for the solutions shown in Fig. \ref{fig:LPDSD0}. Panels b. and c. show cross cuts of the flow velocity. Panel b. plots the $u$ component of the flow for a horizontal cut of the cavity located at $x=0.5$, whereas panel c. shows a vertical cut located at $y=0.5$. The markers indicate values extracted from Table I of Ref. \cite{ghia1982high}, the solid blue curves indicate results from the $\left( \psi,p \right)$ formulation, and the blue dashed curves indicate results from the $\left( v,p\right)$ formulation. 15,000 iterations with ADAM and 200,000 iterations with L-BFGS were performed.}
\label{fig:LPDSD2}
\end{figure}

\begin{figure}
\begin{centering}
\subfigure[]{\includegraphics[scale=0.5]{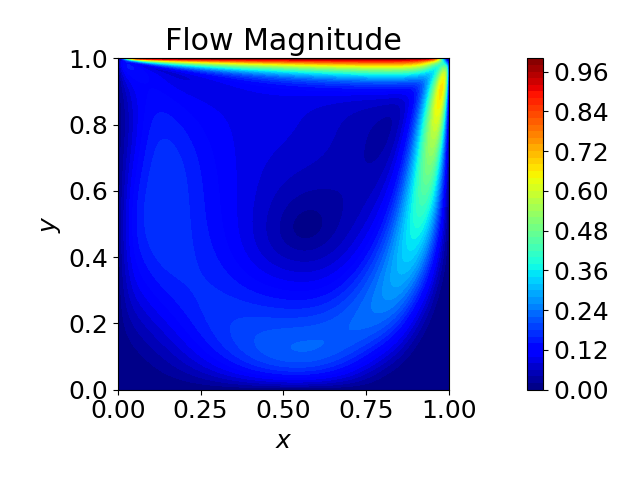}}
\subfigure[]{\includegraphics[scale=0.5]{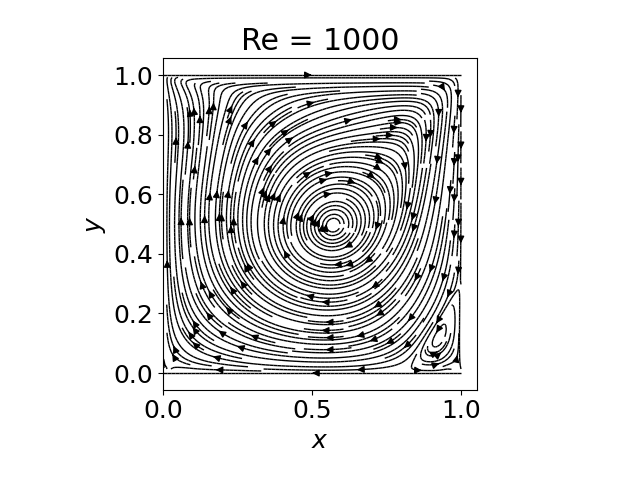}}
\subfigure[]{\includegraphics[scale=0.5]{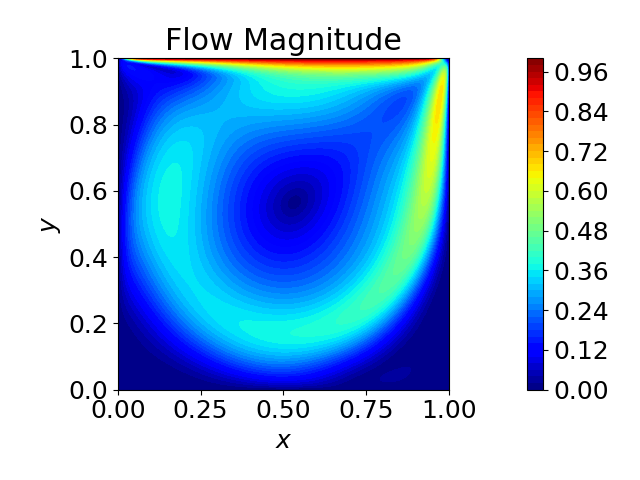}}
\subfigure[]{\includegraphics[scale=0.5]{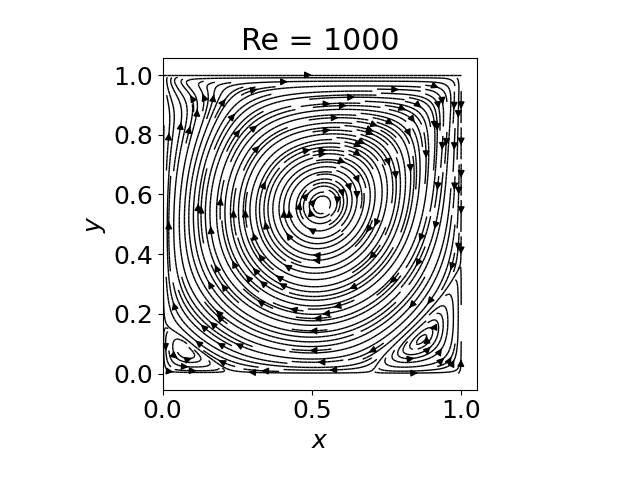}}
\par\end{centering}
\caption{Flow solutions using the $\left( v,p\right)$ (top row) and $\left( \psi, p\right)$ (bottom row) formulations. Flow magnitudes are indicated in the left column and streamlines in the right column. The Reynolds number was taken to be $Re=1000$ and the cavity was taken to have an aspect ratio of $K=1$. A NN with an architecture of $(3) + (32) \times 4 + (3)$ was used, 50,000 training points were employed, with 15,000 iteration with ADAM and 200,000 iterations with L-BFGS.}
\label{fig:LPDSD1}
\end{figure}

Example solutions using these approaches are shown in Figs. \ref{fig:LPDSD0} and \ref{fig:LPDSD1} for Reynolds numbers of $Re=400$ and $Re=1000$, respectively. While in subsequent sections of the paper will will train models to learn the parameter space of the lid driven cavity spanned by $\left(x,y,Re,K, a\right)$, here to demonstrate the two approaches the input to the model will be $x,y$ for a specific Reynolds number and we assume a square cavity. We note that the case of a square cavity was previously treated using the PINN, and conservative PINN (cPINN) methods in Refs. \cite{jagtap2020conservative, amalinadhi2022physics} for a Reynolds number of $Re=100$. For the lower Reynolds number case, both approaches are able to capture the essential aspects of the flow profiles (see Fig. \ref{fig:LPDSD2}). Specifically, a circulation pattern (i.e. primary vortex) is driven throughout the cavity, where the magnitude of the $u$ and $v$ components of the computed flows are in good agreement with the well established results of Ref. \cite{ghia1982high}. More subtlety, secondary vortices are also expected to be present in the bottom corners of the cavity for both of the Reynolds numbers considered here. These secondary vortices have flow magnitudes that are orders of magnitude smaller than the primary vortex, and thus pose a challenge for PINNs, since their contribution to the overall loss function is dwarfed by the contribution from the primary vortex. Nevertheless, both approaches are able to recover the presence of these secondary vortices for the $Re=400$ case. For the higher Reynolds number case of $Re=1000$, the $\left( v, p\right)$ formulation struggles to capture qualitative aspects of the primary and secondary vortices. Specifically, while high resolution simulations~\cite{ghia1982high} have shown that the area occupied by the secondary vortices increases when moving from the $Re=400$ to $Re=1000$ case, the $\left( v, p\right)$ formulation predicts the bottom left secondary vortex to disappear. The $\left( \psi, p\right)$ formulation, in contrast, correctly predicts the increase of both secondary vortices.


\begin{figure}
\begin{centering}
\subfigure[]{\includegraphics[scale=0.33]{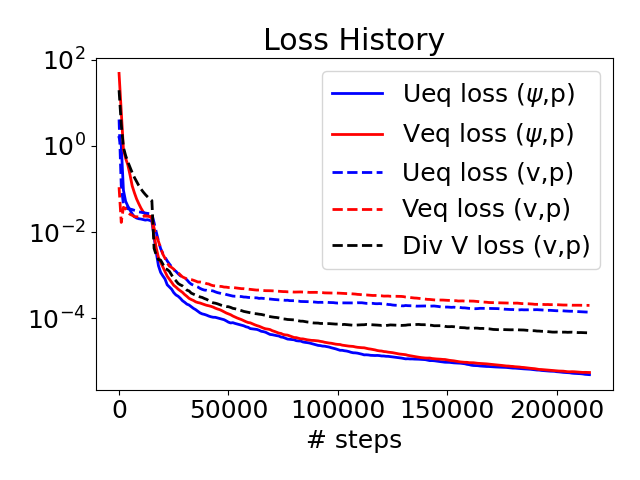}}
\subfigure[]{\includegraphics[scale=0.33]{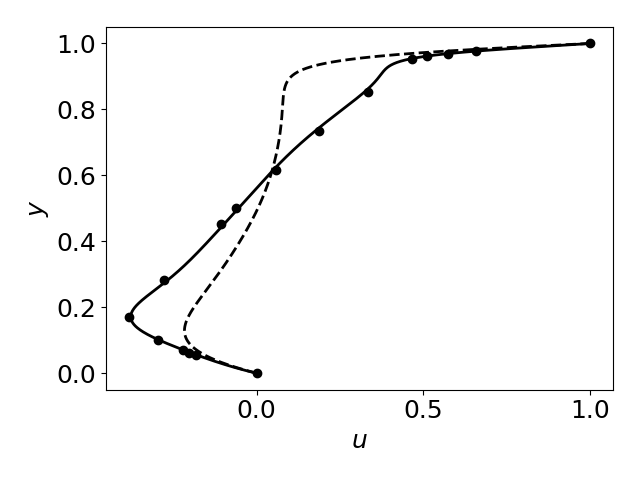}}
\subfigure[]{\includegraphics[scale=0.33]{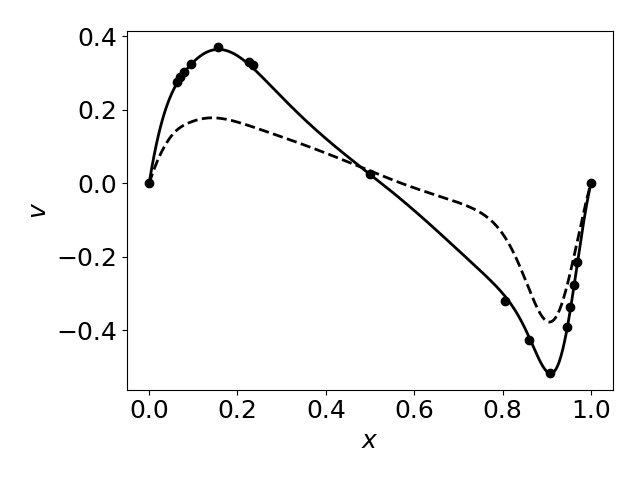}}
\par\end{centering}
\caption{Training loss for the solutions shown in Fig. \ref{fig:LPDSD1}. Panels b. and c. show cross cuts of the flow velocity. Panel b. plots the $u$ component of the flow for a horizontal cut of the cavity located at $x=0.5$, whereas panel c. shows a vertical cut located at $y=0.5$. The markers indicate values extracted from Table I of Ref. \cite{ghia1982high}, the solid black curves indicate results from the $\left( \psi,p \right)$ formulation, and the black dashed curves indicate results from the $\left( v,p\right)$ formulation. 15,000 iterations with ADAM and 200,000 iterations with L-BFGS were performed.}
\label{fig:LPDSD2sub1}
\end{figure}

More quantitative comparisons between the two approaches are shown in Figs. \ref{fig:LPDSD2} and \ref{fig:LPDSD2sub1}. First considering the loss histories of the two approaches, it is apparent that while both models are able to achieve relatively low values for the loss functions, the $\left( \psi, p\right)$ approach is lower for both Reynolds numbers considered. In particular, for the $Re=1000$ case (Fig. \ref{fig:LPDSD2sub1}), the $\left( v, p\right)$ approach is able to achieve a loss of approximately $10^{-4}$, whereas the $\left( \psi, p\right)$ formulation achieves a value of approximately $5\times 10^{-6}$ for the same hyper parameters. We also note that the loss for the $\left( \psi, p\right)$ case is still decreasing implying more iterations would yield a significantly more accurate result. Comparing now the magnitudes of the predicted flows with the well established results from Ref. \cite{ghia1982high}, it is apparent that while both approaches are able to accurately describe the flow for the $Re=400$ case (see Fig. \ref{fig:LPDSD2}), the $\left( v,p\right)$ formulation fails to accurately recover the flow solution for the $Re=1000$ case (see Fig. \ref{fig:LPDSD2sub1}). The $\left( \psi, p\right)$ formulation, in contrast, gives a quantitatively accurate solution for $Re=1000$. While improved results using the $\left( v, p\right)$ formulation can be achieved by further tuning hyper parameters, the apples-to-apples comparison described here suggests that a substantial improvement in accuracy can be achieved at high Reynolds numbers by exactly enforcing incompressibility. 


\section{\label{sec:SC}Square Cavity}

In this section we will consider the relatively simple case of a square cavity. For this case we will seek to learn the solution in the three dimensional space given by $\left(x, y, Re \right)$. We first will explore to what extent a PINN is capable of describing flow and pressure profiles in a square cavity in the absence of data. This approach will fail at sufficiently high Reynolds numbers, and we will thus utilize a small quantity of flow data from high resolution CFD simulations to extend the range of Reynolds numbers that can be accurately described.

\subsection{\label{sec:SCZDL}Zero Data Limit}

\begin{figure}
\begin{centering}
\subfigure[]{\includegraphics[scale=0.5]{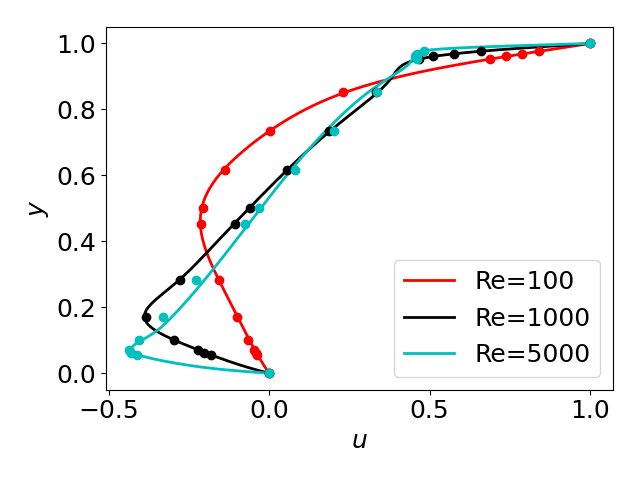}}
\subfigure[]{\includegraphics[scale=0.5]{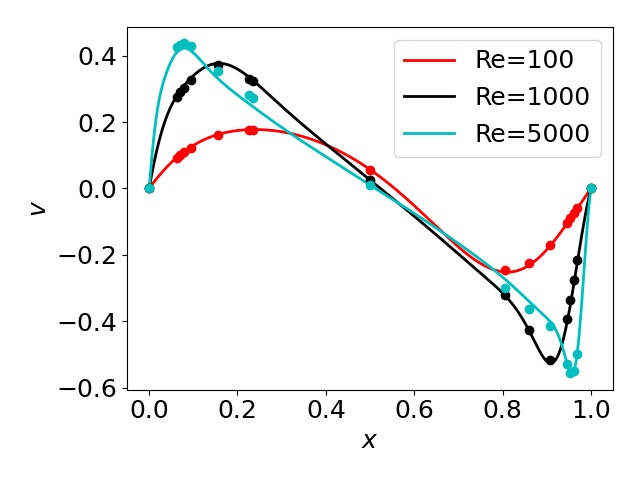}}
\par\end{centering}
\caption{Panels (a) and (b) show comparisons with Ref. \cite{ghia1982high} for three different Reynolds numbers. Panel (b) indicates a vertical cut at $x=0.5$, whereas panel (c) is a horizontal cut at $y=0.5$. The solid curves are the present results, whereas the markers are from Tables I and II of Ghia et al. \cite{ghia1982high}. A NN with an architecture of $(3) + (64) \times 6 + (3)$ was used with 250,000 training points were employed. Roughly 800,000 iterations were performed, primarily using the L-BFGS optimizer.}
\label{fig:LPDSDC2sub1}
\end{figure}

\begin{figure}
\begin{centering}
\subfigure[]{\includegraphics[scale=0.5]{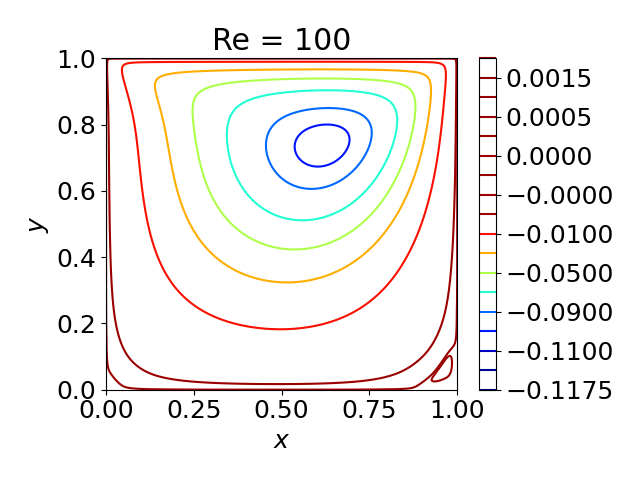}}
\subfigure[]{\includegraphics[scale=0.5]{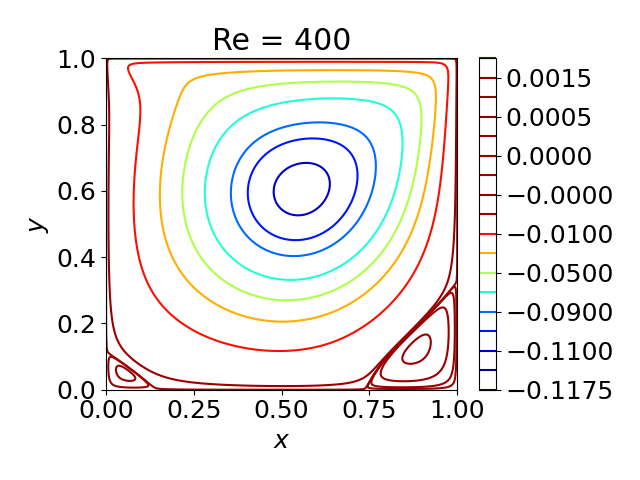}}
\subfigure[]{\includegraphics[scale=0.5]{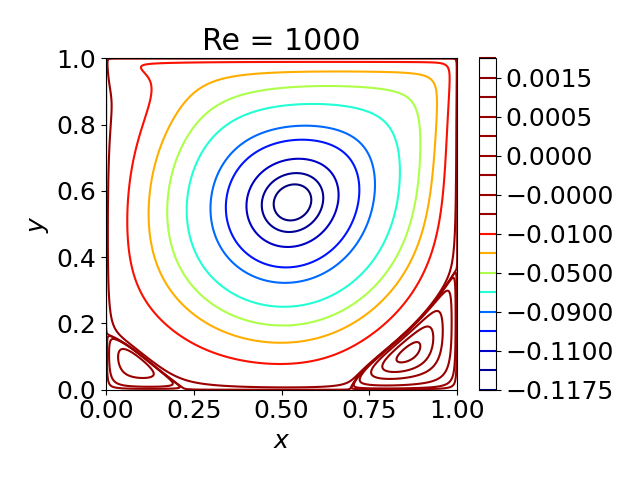}}
\subfigure[]{\includegraphics[scale=0.5]{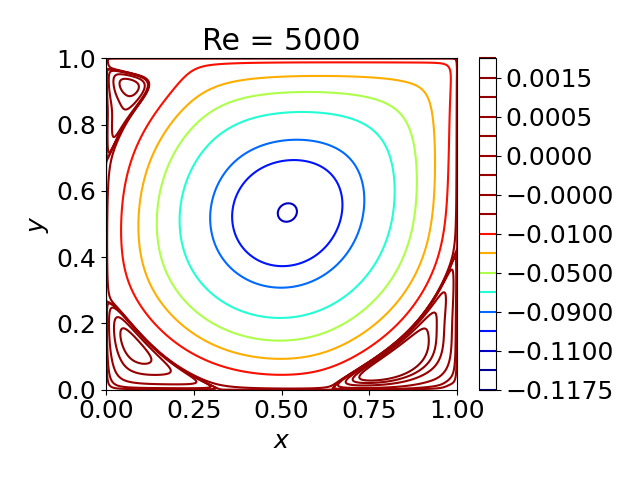}}
\par\end{centering}
\caption{Stream function for different Reynolds numbers computed using the $\left( \psi, p\right)$ formalism. The other parameters are the same as Fig. \ref{fig:LPDSDC2sub1}.}
\label{fig:LPDSDC2sub2}
\end{figure}

It will be of interest to begin by determining the ability of the physics-constrained NN to infer the flow solution in the zero data limit. Noting that the $\left( \psi, p \right)$ formulation provided substantially improved accuracy for $Re \sim 1000$, we will focus exclusively on this approach for the remainder of this paper. First considering a model with training points in the range $Re=\left( 100, 5000\right)$, cross cuts of the flow through the center of the cavity are shown in Fig. \ref{fig:LPDSDC2sub1} for three different Reynolds numbers.
Here it is apparent that the model is able to accurately capture the flows up to Reynolds numbers of several thousand, but begins to show quantitative differences around $Re \approx 5000$. Turning to the form of the stream function topology, Fig. \ref{fig:LPDSDC2sub2} shows contours of the stream function for several different Reynolds numbers. At the lowest Reynolds number considered ($Re=100$), a primary vortex dominates nearly all of the cavity with a very small secondary vortex present in the bottom right corner. As the Reynolds number is increased, the secondary vortex in the bottom right corner grows, and an additional secondary vortex in the bottom left corner forms. Further, increasing the Reynolds number leads to these two secondary vortices increasing in size, with the formation of a third secondary vortex in the top left corner occurring for the $Re=5000$ case. The above trends are qualitatively consistent with high resolution CFD simulations~\cite{ghia1982high, erturk2005numerical}.

\begin{figure}
\begin{centering}
\subfigure[]{\includegraphics[scale=0.5]{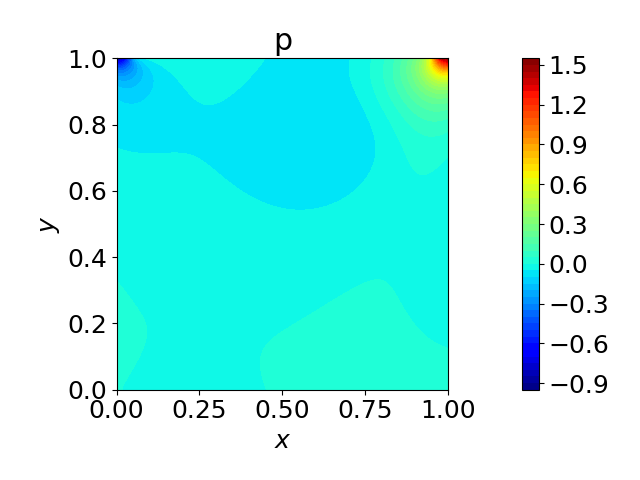}}
\subfigure[]{\includegraphics[scale=0.5]{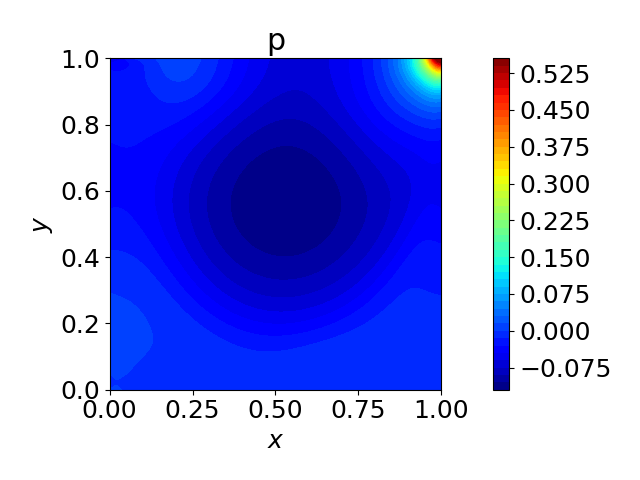}}
\par\end{centering}
\caption{Pressure profile for $Re=100$ [panel (a)] and $Re=1000$ [panel (b)]. The other parameters are the same as Fig. \ref{fig:LPDSDC2sub1}.}
\label{fig:LPDSDC2sub2sub1}
\end{figure}

Aside from the stream function and flow velocities, the pressure is also evaluated in the present approach. Two example pressure profiles are shown in Fig. \ref{fig:LPDSDC2sub2sub1} for both a low and high Reynolds number case. For the low Reynolds number case ($Re=100$), a strong peak in the pressure gradient is evident in the top right corner of the simulation domain, with a corresponding low pressure region located in the top left corner. As the Reynolds number is increased ($Re=1000$) the sharp increase in pressure in the top right corner remains, however the minimum in pressure is instead located at the center of the primary vortex.

\begin{figure}
\begin{centering}
\subfigure[]{\includegraphics[scale=0.33]{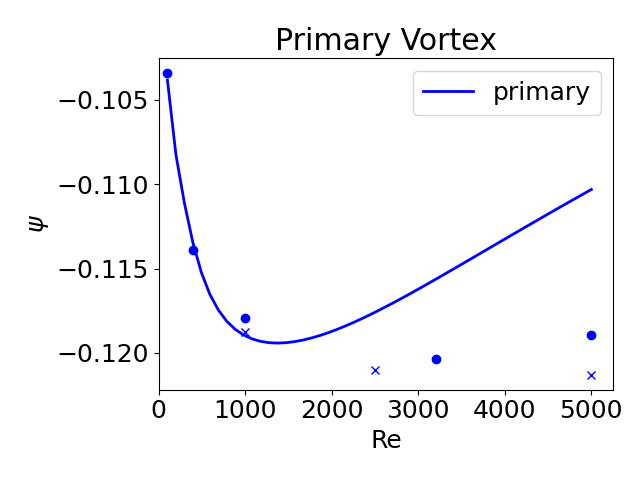}}
\subfigure[]{\includegraphics[scale=0.33]{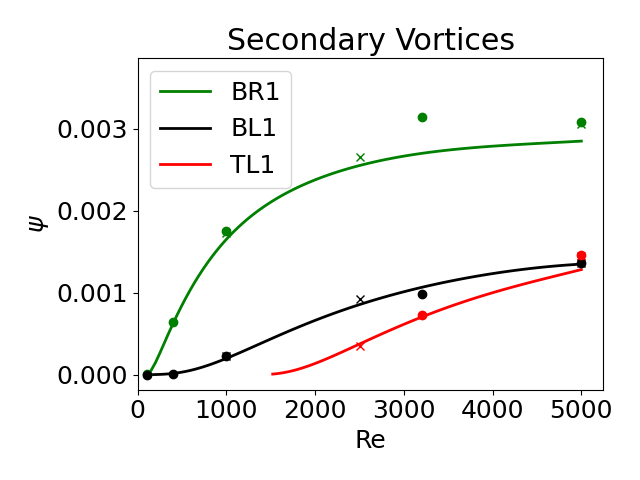}}
\subfigure[]{\includegraphics[scale=0.33]{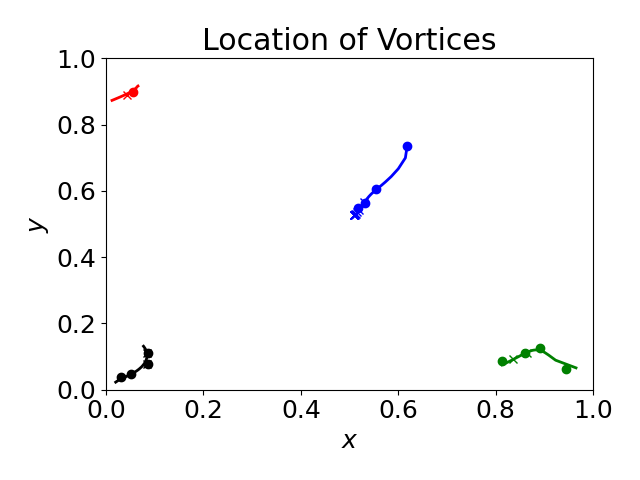}}
\par\end{centering}
\caption{Properties of the primary [panel (a)] and secondary [panel (b)] vortices as a function of Reynolds number. The location of the primary and secondary vortices as the Reynolds number is scanned is shown in panel (c). The `x' markers indicate comparisons with Ref. \cite{erturk2005numerical} whereas the solid `o' markers represent data from Ref. \cite{ghia1982high}. The other parameters are the same as Fig. \ref{fig:LPDSDC2sub1}.}
\label{fig:LPDSDC2sub3}
\end{figure}

A more quantitative comparison with the location and magnitude of different vortices is illustrated in Fig. \ref{fig:LPDSDC2sub3}. Here, the magnitude and location of the four vortices described above as a function of Reynolds number are compared with previous CFD simulations. Focusing on the primary vortex first, the magnitude of the stream function at the center of the primary vortex is in excellent agreement with high resolution CFD simulations~\cite{ghia1982high, erturk2005numerical} for Reynolds number of $Re\lesssim 1000$. As the Reynolds number is further increased to a value between $Re=1000-5000$, quantitative differences begin to emerge. This is not surprising in light of the modest deviations of the flow velocity evident in Fig. \ref{fig:LPDSDC2sub1} for the $Re=5000$ case. Turning now to the secondary vortices, the magnitude of these vortices are approximately two orders of magnitude smaller than the primary vortex suggesting that predicting their formation and evolution will pose a greater challenge to the NN. Somewhat counter intuitively, however, the NN is able to accurately predict the formation and magnitude of all three secondary vortices for the entire range of Reynolds numbers the NN was trained over. Specifically, the vortices in the bottom corners, denoted by `BL1' and `BR1' for bottom left and bottom right, respectively, are present for all Reynolds numbers considered and the amplitude increases with Reynolds number at a rate consistent with predictions from Refs. \cite{ghia1982high, erturk2005numerical}, where the largest deviation is for the BR1 vortex at high Reynolds number. Noting that the results from Refs. \cite{ghia1982high, erturk2005numerical} also exhibit some scatter at these Reynolds numbers implies this region is more challenging to resolve. Turning to the secondary vortex in the top left corner, which we denote by `TL1,' this vortex is predicted to form at a Reynolds number of $Re\approx1500$ and grow as the Reynolds number is increased. While the authors are unaware of any data identifying the critical Reynolds number for the formation of this vortex, $Re\approx1500$ is consistent with the datasets provided in Refs. \cite{ghia1982high, erturk2005numerical}. Considering now the location of the vortices as a function of Reynolds number, Fig. \ref{fig:LPDSDC2sub3}(c) compares results from the literature (`x' and `o' markers) with the results of the present surrogate model (solid curves). It is evident that the present model is able to accurately identify the location of both primary and secondary vortices as the Reynolds number is varied.

\subsection{\label{sec:SDCSC}Small Data Case}

\begin{figure}
\begin{centering}
\subfigure[]{\includegraphics[scale=0.5]{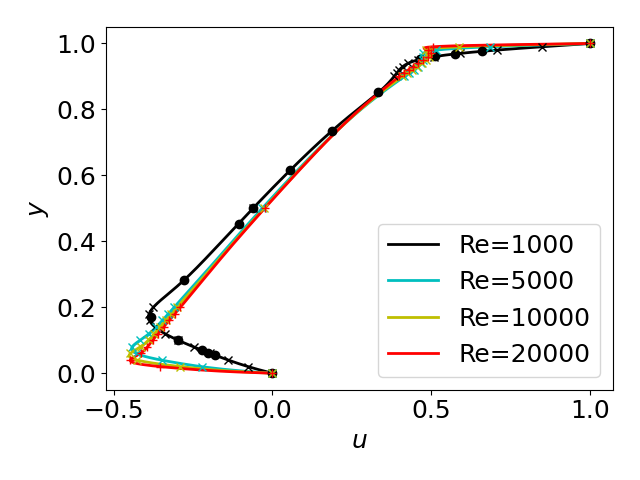}}
\subfigure[]{\includegraphics[scale=0.5]{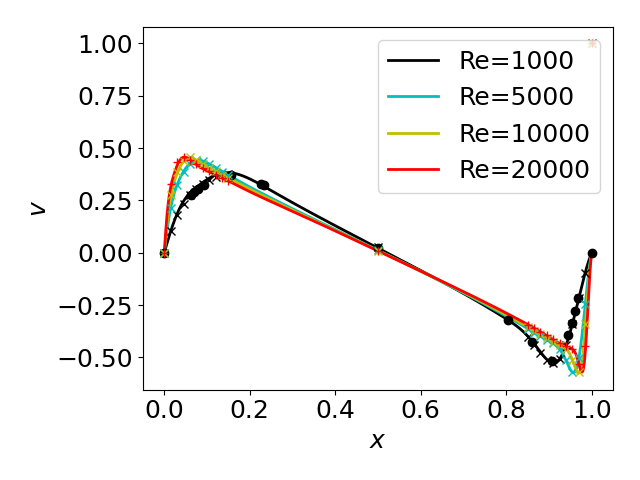}}
\par\end{centering}
\caption{Panels (a) and (b) show comparisons with Refs. \cite{erturk2005numerical} (`x' markers) and \cite{ghia1982high} (solid circles) for several Reynolds numbers. Panel (b) indicates a vertical cut at $x=0.5$, whereas panel (c) is a horizontal cut at $y=0.5$. The solid curves are the present results, whereas the markers are from Tables I and II of Ref. \cite{ghia1982high}. A NN with an architecture of $(3) + (64) \times 6 + (3)$ was used, 250,000 training points were employed, with 15,000 iterations with ADAM, and 285,000 with L-BFGS. A learning rate of $5\times 10^{-4}$ was used.}
\label{fig:LPDSDC3}
\end{figure}

In this section, we will use a small amount of data to aid in inferring the flow and pressure. The flow data used for training the model will be limited to data given in Table VI of Ref. \cite{erturk2005numerical}. This data set only provides the $u$ component of the velocity at fifteen spatial locations, for ten different Reynolds numbers ($Re=1000,2500,5000,7500,10000,7500,10000,12500,15000,17500,20000,21000$). The spatial locations where the $u$ component of the flow is provided corresponds to a vertical line located at $x=0.5$. No information on the $v$ or $p$ profiles will be used, thus this limited data set represents an extremely incomplete picture of the global flow and pressure profiles. We will thus rely heavily on the physics constraints in the loss function to infer the variation of the flow and pressure as a function of both space and Reynolds number.

\begin{figure}
\begin{centering}
\subfigure[]{\includegraphics[scale=0.5]{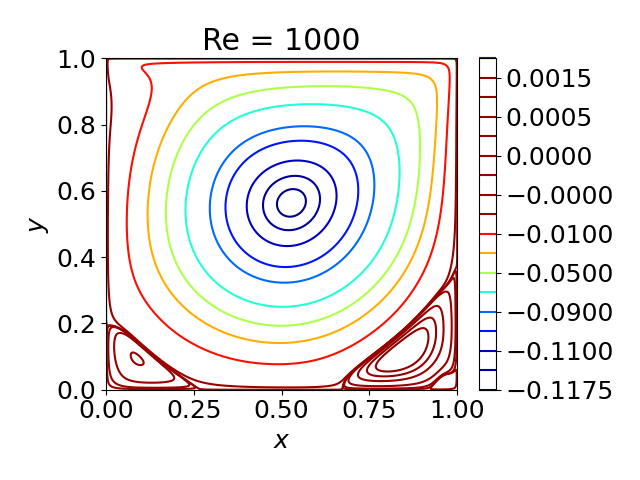}}
\subfigure[]{\includegraphics[scale=0.5]{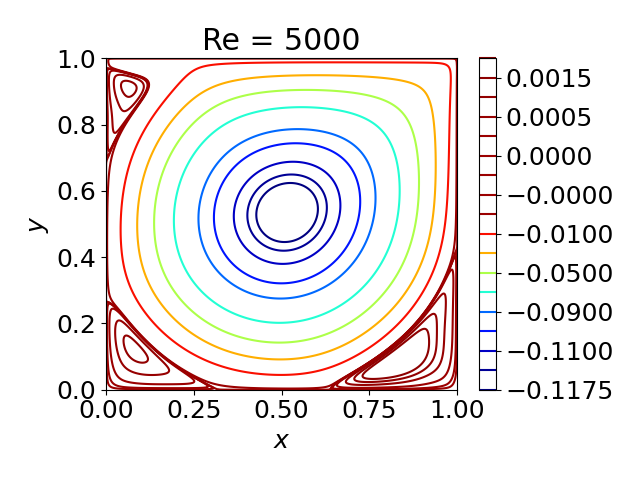}}
\subfigure[]{\includegraphics[scale=0.5]{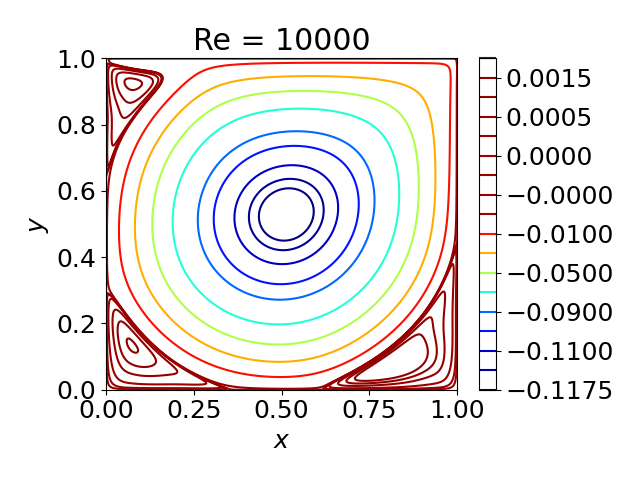}}
\subfigure[]{\includegraphics[scale=0.5]{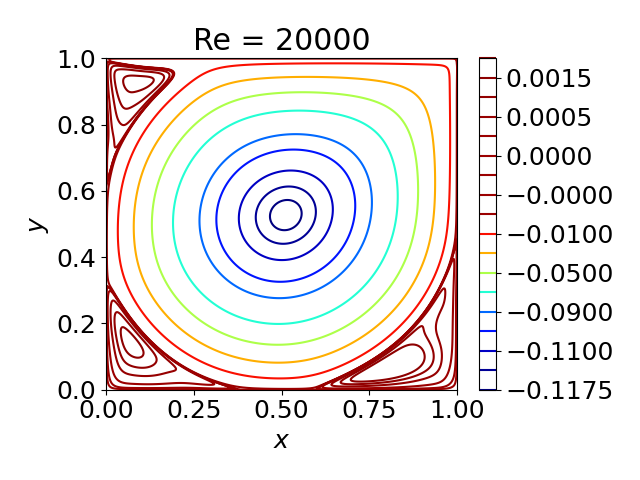}}
\par\end{centering}
\caption{Flow streamlines for different Reynolds numbers. The other parameters are the same as Fig. \ref{fig:LPDSDC3}.}
\label{fig:LPDSDC4}
\end{figure}

Carrying out an analogous calculation as that described in Sec. \ref{sec:SCZDL} above, but using the small data set described above, yields the flows shown in Fig. \ref{fig:LPDSDC3}. Here we have modified the range of Reynolds numbers considered to span $Re=\left( 1000,21000\right)$. It is evident that the $u$ component of the flow is in excellent agreement with Ref. \cite{erturk2005numerical} for the full range of Reynolds numbers scanned [$Re=\left( 1000,21000\right)$]. Such excellent agreement is expected since this data was used when training the network. More nontrivially, the $v$ component of the flow measured at a horizontal line at $y=0.5$ is also in excellent agreement. Data for this component of the flow was not incorporated into the training of the NN, implying the physics constraint in the loss function is able to reconstruct this flow component. 

A more challenging test of this approach will be to investigate how the flow topology changes as the Reynolds number is increased. This is shown in Fig. \ref{fig:LPDSDC4}. Here it is apparent that the qualitative flow structures are reproduced up to a Reynolds number of $Re\approx 5000$. Specifically, for the lowest Reynolds number considered ($Re=1000$), secondary vortices are present in the bottom corners.  As the Reynolds number is further increased ($Re=5000$) an additional secondary vortex appears in the top left corner. However, as the Reynolds number is increased to $Re=10,000$, high resolution CFD simulations suggest tertiary vortices form in the bottom corners of~\cite{ghia1982high,erturk2005numerical}. These vortices are not evident in Fig. \ref{fig:LPDSDC4}(c), or at a higher Reynolds number of $Re=20,000$ [Fig. \ref{fig:LPDSDC4}(d)], where the tertiary vortices are expected to be more pronounced. Additionally, at $Re=20,000$ an additional tertiary vortex is expected to form in the top left corner~\cite{erturk2005numerical}, which is not evident in Fig. \ref{fig:LPDSDC4}(d). The challenge in recovering these tertiary vortices is likely due to their small magnitude. While secondary vortices tend to have a magnitude that is roughly two orders of magnitude smaller than the primary vortex, tertiary vortices are three orders of magnitude smaller. Their accurate resolution thus poses a challenge for the PINN to resolve. We note in passing, that with additional training, or varying various hyperparameters of the NN, the formation of a tertiary vortex in the bottom corners can sometimes be reproduced, however, their appearance is not robustly recovered by the current approach.

\begin{figure}
\begin{centering}
\subfigure[]{\includegraphics[scale=0.33]{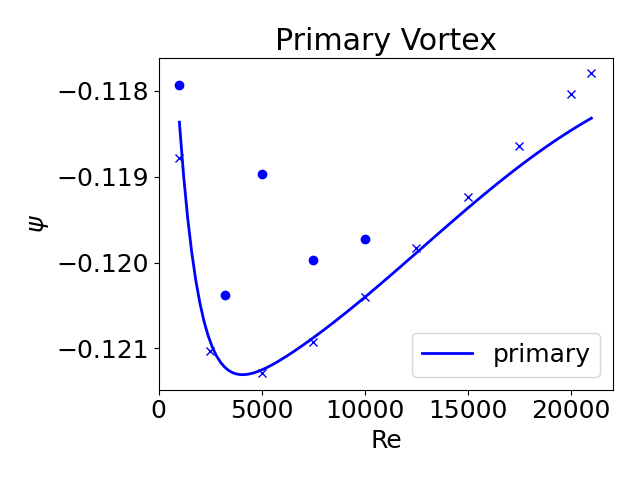}}
\subfigure[]{\includegraphics[scale=0.33]{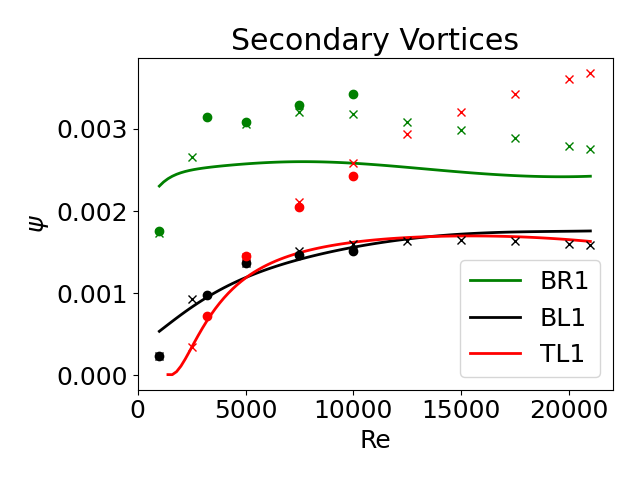}}
\subfigure[]{\includegraphics[scale=0.33]{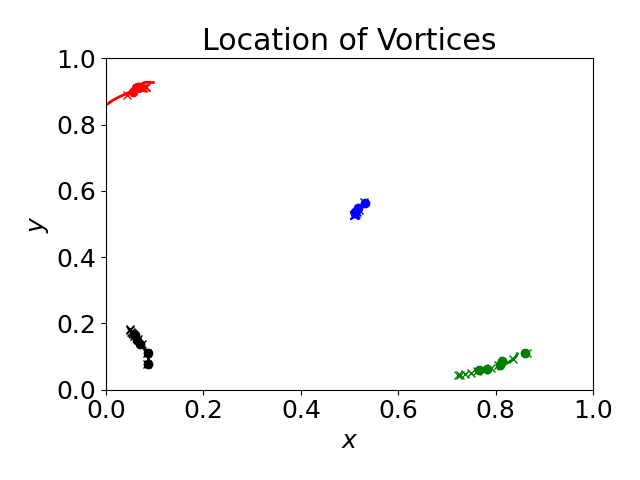}}
\par\end{centering}
\caption{Properties of the primary [panel (a)] and secondary [panel (b)] vortices as a function of Reynolds number. The location of the primary and secondary vortices as the Reynolds number is scanned is shown in panel (c). The `x' markers indicate comparisons with Ref. \cite{erturk2005numerical} whereas the solid `o' markers represent data from Ref. \cite{ghia1982high}. A NN with an architecture of $(3) + (64) \times 6 + (3)$ was used, 250,000 training points were employed, with 15,000 iterations with ADAM, and 285,000 with L-BFGS. A learning rate of $5\times 10^{-4}$ was used.}
\label{fig:LPDSDC3sub1}
\end{figure}

Quantitative comparisons of the location, and magnitude of the primary vortex and the three secondary vortices are shown in Fig. \ref{fig:LPDSDC3sub1}. Considering first the primary vortex [Fig. \ref{fig:LPDSDC3sub1}(a)], the variation of its magnitude as a function of Reynolds number is accurately reproduced for nearly all Reynolds numbers considered, with only a slight deviation for the highest Reynolds numbers near $Re\sim 20,000$. In particular, for Reynolds numbers between $Re=\left( 1000, 10000\right)$ the present results appear to be in better agreement with Ref. \cite{erturk2005numerical} compared to the classic study of Ref. \cite{ghia1982high}. Turning to the secondary vortices, good agreement with regard to their magnitude is present for Reynolds numbers between $Re=\left( 1000,5000\right)$, however, for larger Reynolds numbers the present results deviate from Ref. \cite{erturk2005numerical}. This is not unexpected, since it is in this range of Reynolds numbers that tertiary vortices are expected to form, which are not recovered in the present approach. Additional training can modestly improve agreement, though even after training for over a million iterations, the NN is unable to accurately describe the magnitude of the secondary vortices for the full range of Reynolds numbers.

\section{\label{sec:TC}Trapezoidal Cavity}

It will now be of interest to consider a more challenging problem, whereby aside from varying the Reynolds number, we will seek to learn how the flow varies as a function of the cavity geometry. For definiteness we will choose a trapezoidal geometry, which is parameterized by the normalized depth of the cavity $K\equiv D/L$ and the offset of the lower vertices from $x=0$ and $x=L$ is parameterized by the parameter $a\equiv \delta / L$ (see Fig. \ref{fig:Cavity}). In the limit $a\to0$ a rectangular cavity will be recovered, whereas for $a\to 0.5$ a triangular geometry will be present. The surrogate model described here will thus attempt to recover the flow and pressure profile for a five-dimensional space. We will begin by considering the zero data limit and determine the ability of the PINN to recover aspects of the flow and pressure profile as a function of Reynolds number and cavity geometry. Once we have quantified the accuracy of the zero data limit, we will utilize a small amount of data from CFD simulations to improve the accuracy of the surrogate model.

\subsection{\label{sec:TCZDL}Zero Data Limit}

\begin{figure}
\begin{centering}
\subfigure[]{\includegraphics[scale=0.5]{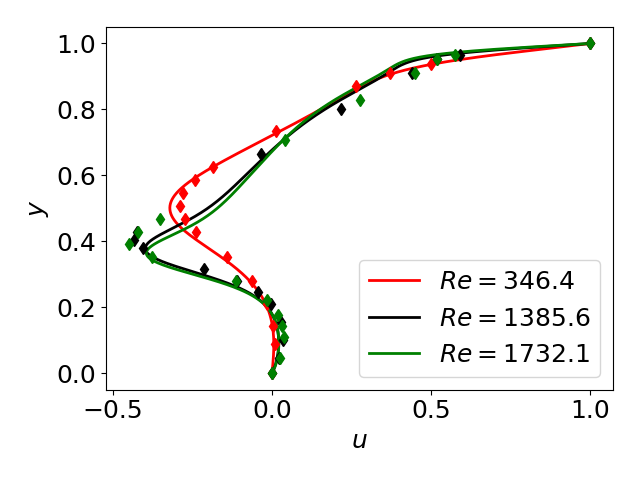}}
\subfigure[]{\includegraphics[scale=0.5]{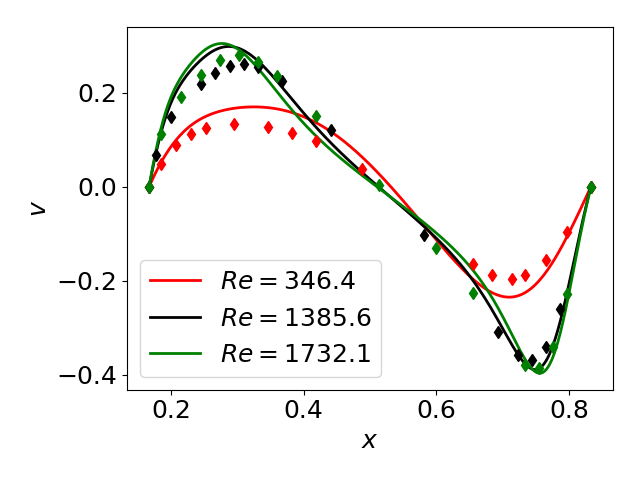}}
\subfigure[]{\includegraphics[scale=0.5]{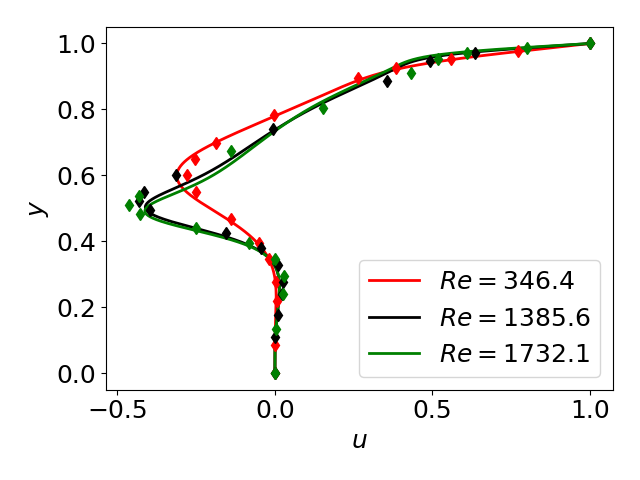}}
\subfigure[]{\includegraphics[scale=0.5]{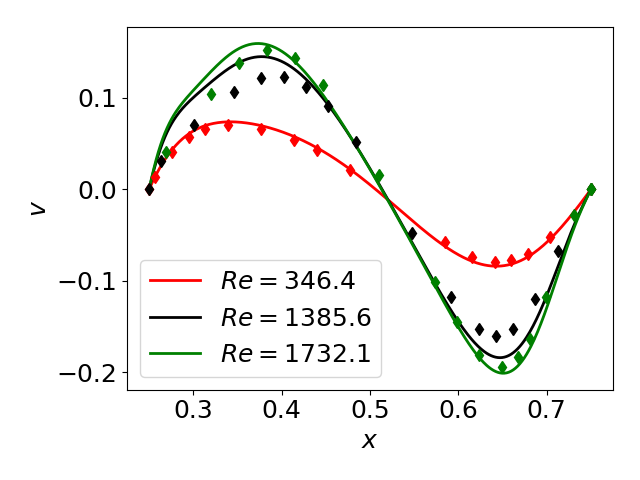}}
\par\end{centering}
\caption{Comparison with the velocity profiles computed in Ref. \cite{paramane2008consistent} in the absence of flow data. Panels (a) and (b) indicate the predicted flow for $a=1/3$, whereas panels (c) and (d) indicate the predicted flow for $a = 1/2$ (triangular cavity). The solid curves correspond to predictions from the present model, whereas the diamond markers indicate results from Ref. \cite{paramane2008consistent}. A NN with an architecture of $(5) + (64) \times 6 + (2)$ was used, 250,000 training points were employed, with 15,000 iterations with ADAM, and 285,000 with L-BFGS. A learning rate of $5\times 10^{-4}$ was used.}
\label{fig:TCZDL1}
\end{figure}


For the present example we will train a model for a Reynolds number in the range $Re=\left( 100, 2000\right)$, $K=\left( 0.5, 1\right)$ and $a = \left(0, 0.5\right)$. After training the model by performing 15,000 iterations with the ADAM optimizer and 285,000 iterations with L-BFGS, a comparison with the predicted flows and those recorded in Ref. \cite{paramane2008consistent} is shown in Fig. \ref{fig:TCZDL1}. Here, two different values of $a$ are considered, where in both cases the solutions are in good agreement with Ref. \cite{paramane2008consistent}.
We note in passing that the values of $v$ shown in Fig. \ref{fig:TCZDL1}(d) deviate somewhat from those given in Ref. \cite{paramane2008consistent}, where our results appear to be in closer agreement with the more recent results shown in Fig. 3 of Ref. \cite{zhang2010lattice}. While the present zero data approach provides acceptable agreement with grid based high resolution solvers, and further training would likely improve agreement, it will be informative to compare the present results with those in which a small quantity of data is incorporated in the training of the network.

\subsection{\label{sec:TCSDL}Small Data Limit}

\begin{figure}
\begin{centering}
\subfigure[]{\includegraphics[scale=0.5]{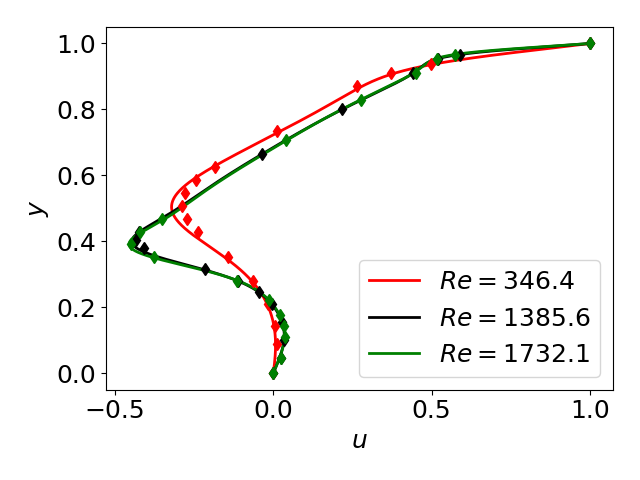}}
\subfigure[]{\includegraphics[scale=0.5]{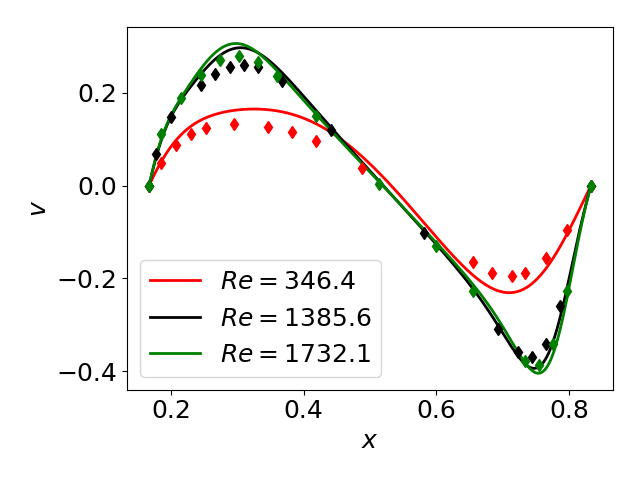}}
\subfigure[]{\includegraphics[scale=0.5]{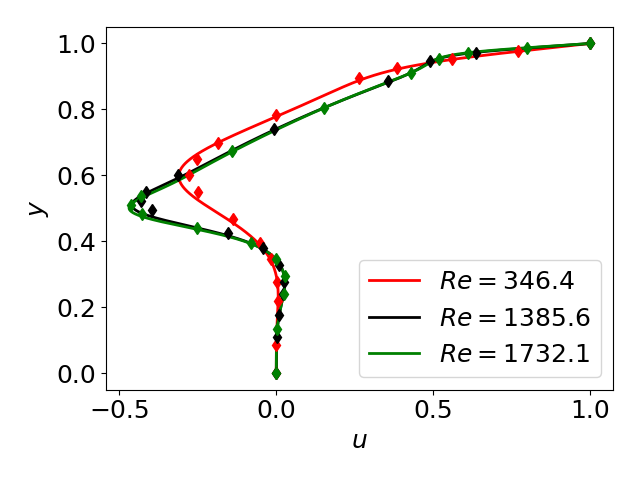}}
\subfigure[]{\includegraphics[scale=0.5]{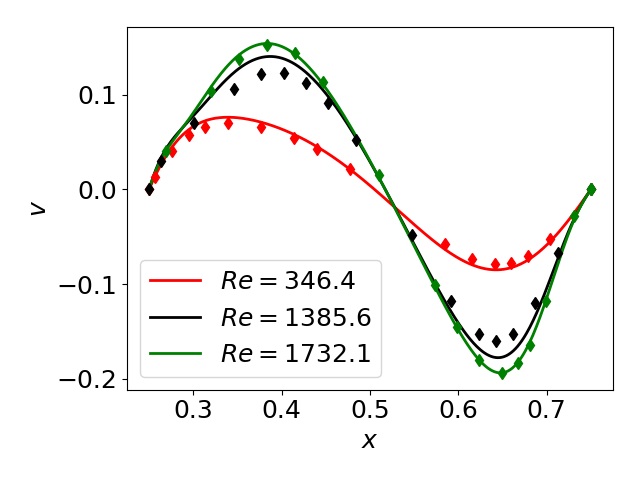}}
\par\end{centering}
\caption{Comparison with the velocity profiles computed in Ref. \cite{paramane2008consistent} using a small quantity of flow data. Panels (a) and (b) indicate the predicted flow for $a=1/3$, whereas panels (c) and (d) indicate the predicted flow for $a = 1/2$ (triangle). The solid curves correspond to predictions from the present model, whereas the diamond markers indicate results from Ref. \cite{paramane2008consistent}. A NN with an architecture of $(5) + (64) \times 6 + (2)$ was used, 250,000 training points were employed, with 15,000 iterations with ADAM, and 285,000 with L-BFGS. A learning rate of $5\times 10^{-4}$ was used.}
\label{fig:TCSDL1}
\end{figure}

The data that will be used to expedite and enhance the accuracy of the NN corresponds to table I of Ref. \cite{ghia1982high}, which applies to the case of a square geometry, and tables 1-3 of Ref. \cite{paramane2008consistent}, which treats trapezoidal and triangular geometry. Once again, we will only use data on the $u$-component of the velocity located at a vertical cut at $x=0.5$. No information about the $v$-component of the flow or the pressure will be used. Fifteen values of the $u$-component of the flow will be used at three different Reynolds numbers. Noting that four different geometries are sampled (square, two trapezoids, and triangular) are used, the total number of values of velocities used in the training corresponds to $15\times 3 \times 4 = 180$. These $180$ measurements will be the only data involved in describing the flow and pressure solutions in the five-dimensional space spanned by $\left( x,y,Re,K,a \right)$. For the present example we will train a model for a Reynolds number in the range $Re=\left( 100,2000\right)$, $K=\left( 0.5, 1\right)$ and $a = \left(0, 0.5\right)$. While the present approach would likely be able to provide accurate treatments at Reynolds numbers far greater than $Re=2000$, we will not treat high Reynolds number cases due to a lack of high resolution published simulation data for such cases. 

As evident by comparing Fig. \ref{fig:TCSDL1} (a case using flow data) with Fig. \ref{fig:TCZDL1} (no data), the incorporation of a small quantity of data leads to substantial improvement in the ability of the NN to reconstruct the global flow profile. An interesting aspect of this comparison, is that more recent high resolution simulations (Fig. 3 of Ref. \cite{zhang2010lattice}) of shear driven flow in trapezoidal geometry have shown that Ref. \cite{paramane2008consistent} underestimates the $v$-component of the flow, while providing an accurate calculation of the $u$-component of the flow. The present results recover this discrepancy, where the predictions of the $v$-component of the flow are consistently higher than those predicted by Ref. \cite{paramane2008consistent}.

\begin{figure}
\begin{centering}
\subfigure[]{\includegraphics[scale=0.5]{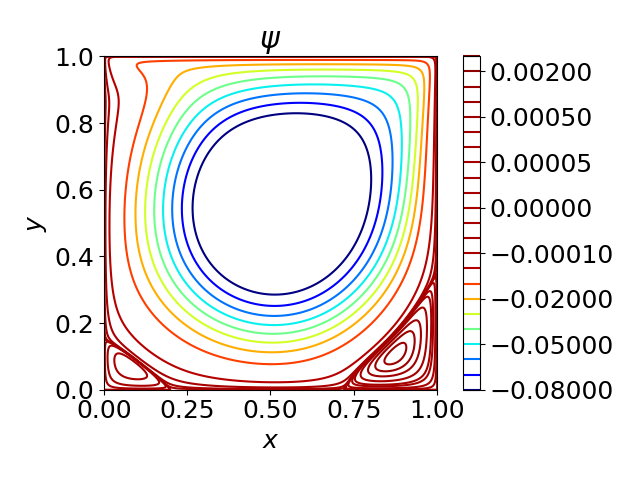}}
\subfigure[]{\includegraphics[scale=0.5]{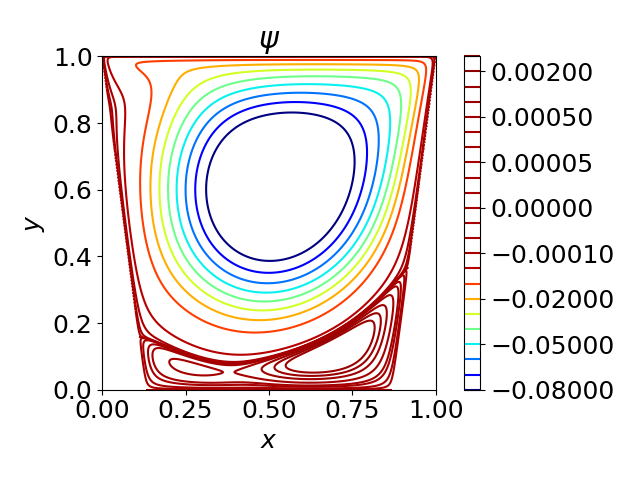}}
\subfigure[]{\includegraphics[scale=0.5]{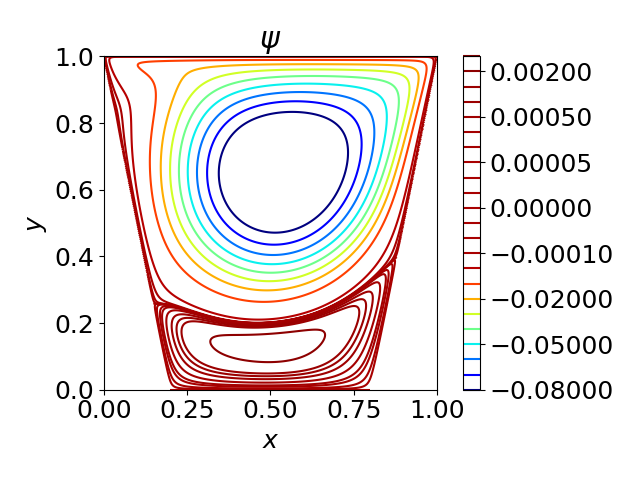}}
\subfigure[]{\includegraphics[scale=0.5]{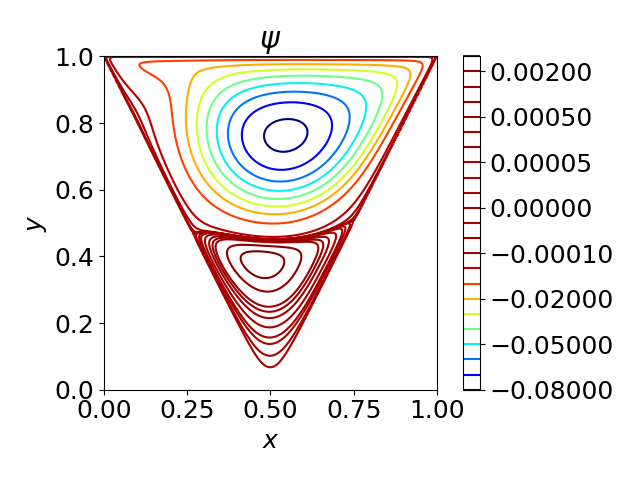}}
\par\end{centering}
\caption{Contours of the stream function before and after vortex merger. Panel (a) is for $a=0$, panel (b) is for $a=0.13$, panel (c) is for $a=0.2$ and panel (d) is for $a=0.5$. The Reynolds number for all cases was taken to be $Re=1000$ and the depth of the cavity was taken to be $K=1$.}
\label{fig:TCSDL2}
\end{figure}

\begin{figure}
\begin{centering}
\subfigure[]{\includegraphics[scale=0.5]{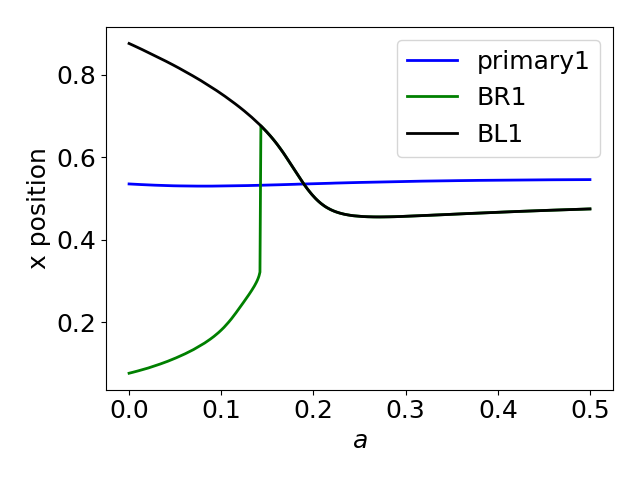}}
\subfigure[]{\includegraphics[scale=0.5]{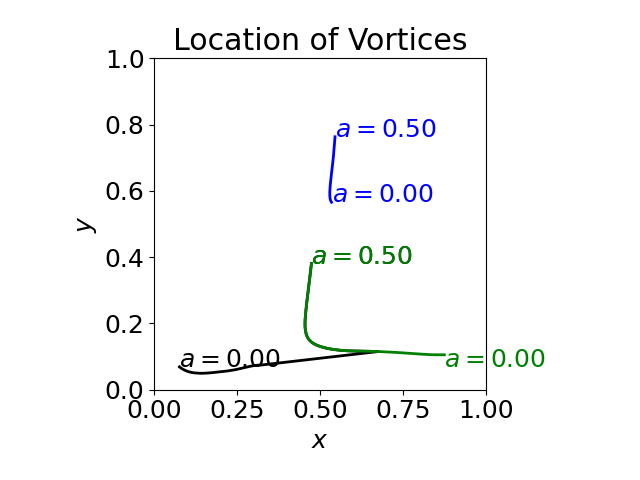}}
\par\end{centering}
\caption{Spatial evolution of the primary and secondary vortices as $a$ is scanned. Panel (a) indicated the $x$-position of the vortices as $a$ is varied. Panel (b) indicates the $\left( x, y \right)$ position of the vortices. The Reynolds number was taken to be $Re=1000$ and the depth of the cavity was taken to be $K=1$.}
\label{fig:TCZDL3}
\end{figure}

An interesting aspect of the trapezoidal cavity problem is that the flow topology undergoes transitions both as the Reynolds number is changed, as well as when the geometry of the cavity is varied. This is illustrated in Fig. \ref{fig:TCSDL2}, where the geometry is varied from a square cavity to a triangular cavity ($a=0.5$) for a Reynolds number of $Re=1000$. As $a$ is increased, the corner vortices approach each other and merge between $a=0.14$ and $a=0.2$. More insight into the evolution of these vortices can be gained by considering how the spatial position of the vortex centers changes as $a$ is scanned. This is shown in Fig. \ref{fig:TCZDL3}, where for $a=0$ two secondary vortices are evident in the bottom corners of the cavity. As $a$ increases these vortices are pushed toward each other and merge at $a\approx 0.14$.

\begin{figure}
\begin{centering}
\subfigure[]{\includegraphics[scale=0.5]{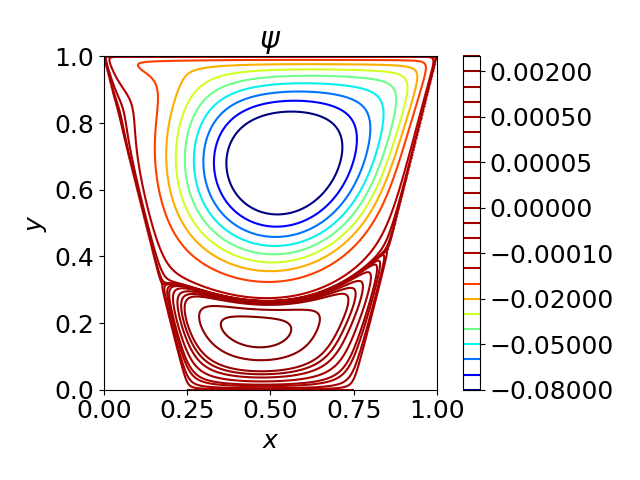}}
\subfigure[]{\includegraphics[scale=0.5]{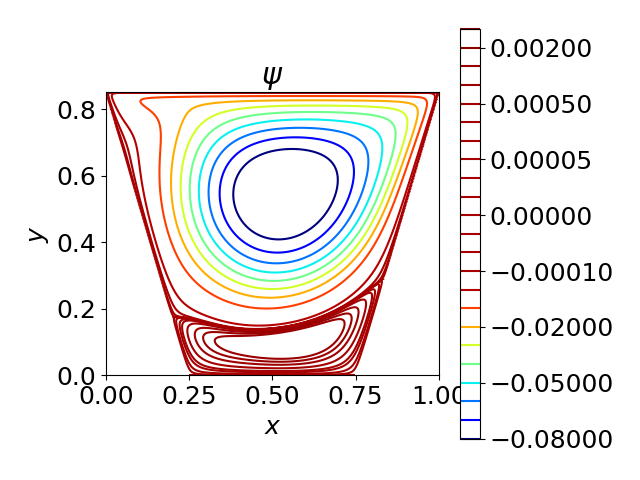}}
\subfigure[]{\includegraphics[scale=0.5]{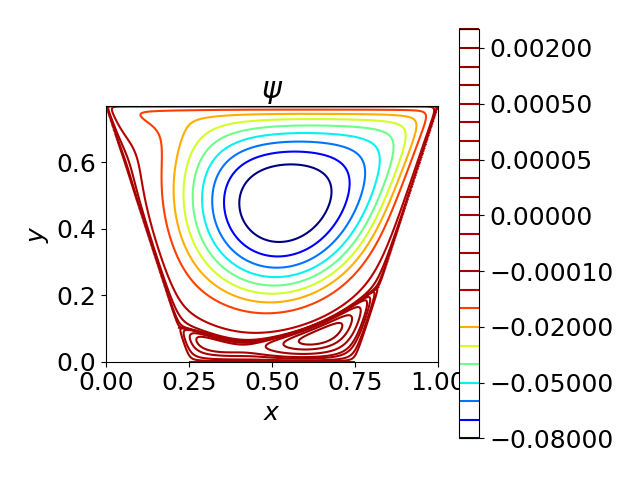}}
\subfigure[]{\includegraphics[scale=0.5]{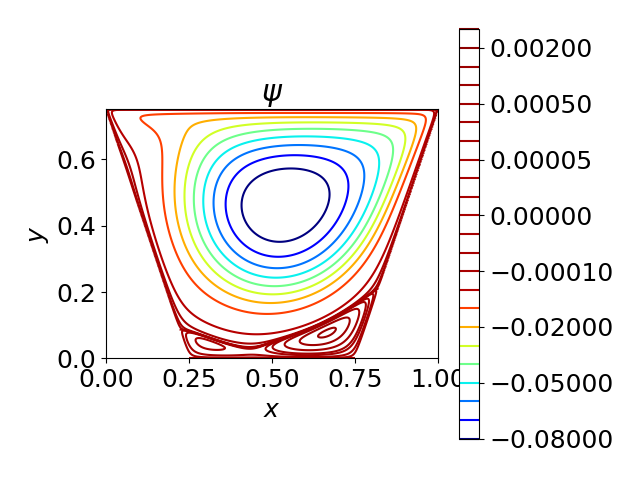}}
\par\end{centering}
\caption{Contours of the stream function before and after vortex splitting. Panel (a) is for $K=1$, panel (b) is for $K=0.85$, panel (c) is for $K=0.77$ and panel (d) is for $K=0.75$. The Reynolds number for all cases was taken to be $Re=1000$ and $a=0.25$.}
\label{fig:TCZDL4}
\end{figure}

\begin{figure}
\begin{centering}
\subfigure[]{\includegraphics[scale=0.5]{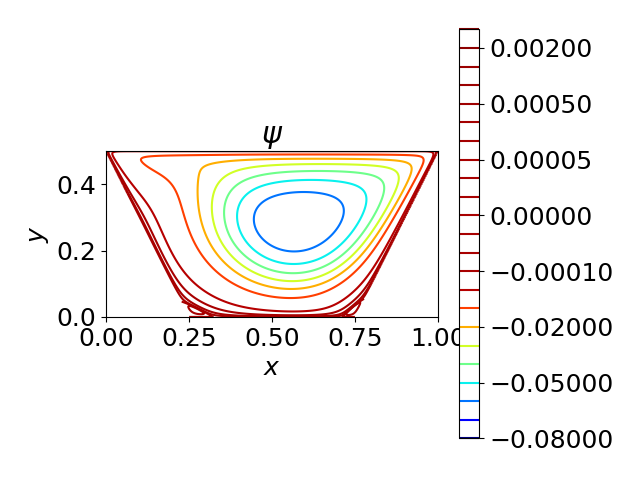}}
\subfigure[]{\includegraphics[scale=0.5]{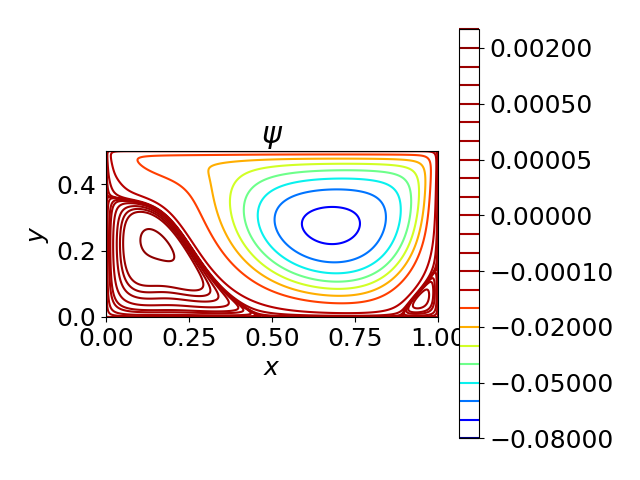}}
\par\end{centering}
\caption{Shrinking and expansion of secondary vortices. Panel (a) is for $K=0.5$ and $a=0.25$ whereas panel (b) is for $K=0.5$ and $a=0$. The Reynolds number for all cases was taken to be $Re=1000$.}
\label{fig:TCZDL5}
\end{figure}

Once the bottom vortex has formed by reducing $a$ it can subsequently be split by shrinking the depth of the cavity while holding $a$ constant. This is illustrated in Fig. \ref{fig:TCZDL4}, where for $a=0.25$ and $K=1$, a single lower vortex is present. However, as the depth $K$ is decreased the primary vortex is forced into the lower vortex, which eventually splits the lower vortex into two secondary vortices for $K \lesssim 0.8$. As the depth of the cavity is further reduced the two bottom secondary vortices will shrink further as they are forced against the primary vortex. In contrast, by decreasing $a$, and thus making the bottom region of the cavity larger, these secondary vortices will expand in size. These two trends are illustrated in Fig. \ref{fig:TCZDL5}, wherein panel (a) the depth of the cavity is reduced to $K=0.5$, resulting in the bottom secondary vortices shrinking in size. Subsequently, as $a\to 0$, resulting in a shallow rectangular cavity, the two bottom vortices increase in magnitude, with a flow structure that recovers that observed in Ref. \cite{cheng2006vortex}.

\begin{figure}
\begin{centering}
\subfigure[]{\includegraphics[scale=0.5]{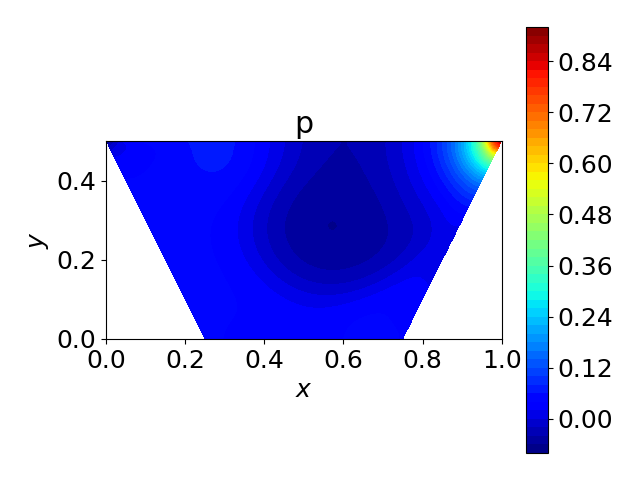}}
\subfigure[]{\includegraphics[scale=0.5]{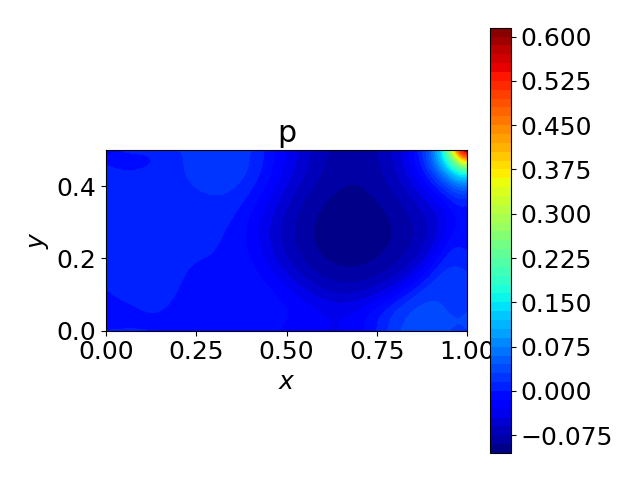}}
\par\end{centering}
\caption{Pressure profiles for a shallow trapezoidal cavity (panel (a), $K=0.5$ and $a=0.25$) and a shallow rectangular cavity (panel (b), $K=0.5$ and $a=0$) and . The Reynolds number for both cases was taken to be $Re=1000$.}
\label{fig:TCZDL6}
\end{figure}

In addition to the changes in the flow structure as the geometry evolves, the pressure profile also undergoes strong variations. Considering the case of a shallow rectangular cavity [see Fig. \ref{fig:TCZDL6}(b)], a high pressure region is present in the top right corner, similar to the case of a square cavity. In contrast, the minimum pressure is offset from the center of the cavity. This is due to the location of the primary vortex also being displaced from the center by the strong secondary vortex in the bottom left corner. As the cavity geometry is changed to a trapezoid, the pressure peak in the top right corner becomes more pronounced, with the low pressure region shifting toward the center of the cavity coinciding with the location of the primary vortex.

\section{\label{sec:C}Discussion and Conclusions}

The present work has utilized deep learning methods to develop a surrogate model of flow in a non-rectangular cavity. While the training of this model is computationally intensive, once trained the surrogate model is able to provide predictions of the stream function, flow velocity and pressure in a few microseconds, thus providing a valuable tool for exploring the five-dimensional space present in a trapezoidal cavity. In the present work this surrogate model was used to identify critical parameters at which changes to the flow topology occur as a function of the Reynolds number and cavity geometry. 


The primary limitation of the PINN approach has been the relative lack of accuracy in comparison to grid based CFD solvers. As shown in Sec. \ref{sec:LDCF}, enforcing boundary conditions and incompressibility as hard constraints substantially increases the accuracy of PINNs for Reynolds numbers $\gtrsim 1000$ for a square cavity. At higher Reynolds numbers ($Re\sim 5000$), the PINN method began to struggle to quantitatively recover the flow profile of the primary vortex. For such high Reynolds number cases, the use of a small quantity of flow data from high resolution CFD simulations allowed the flows associated with the primary vortex to be accurately recovered up to Reynolds numbers of $Re=21,000$. This synergy between the use of data and physical constraints in the training of the NN suggests the intriguing possibility whereby PINNs may act to complement grid based CFD solvers for the purpose of efficiently exploring high dimensional paramters spaces. Specifically, while grid based CFD solvers are capable of providing high resolution descriptions of fluid flows, their application to many-query or real-time analyses is limited due to their computational cost. 
PINNs, while capable of providing predictions in a matter of microseconds, are not able to achieve the same accuracy as grid based CFD solvers in the absence of data. 
The present study has demonstrated that a small quantity of data may be used to extend the range of parameters that PINNs are able to accurately treat, thus enabling the efficient exploration of a high-dimensional parameter space.


While the present work has hinted at the promise of using PINNs as surrogate models for exploring high dimensional parameter spaces, it has also exposed some limitations of the framework. In particular, for the high Reynolds number case explored in Sec. \ref{sec:SDCSC}, while the use of a small quantity of data allowed the PINN to accurately recover the primary vortex, it failed to robustly recover the presence of tertiary vortices. This was due to the relatively small magnitude of these vortices compared to the primary vortex (roughly three orders of magnitude difference). This limitation could potentially be mitigated by additional training, packing training points in the corners of the cavity, or by an adaptive sampling scheme. Alternatively, utilizing high resolution CFD data directly in these regions would provide a direct means of ensuring the resolution of these structures. While the above approaches provide paths toward the resolution of these weak flow structures, it also suggests the present approach cannot be expected to robustly identify subtle properties of complex flows without great care.


\begin{acknowledgments}

This work was supported by the University of Florida Informatics Institute seed award. The authors acknowledge University of Florida Research Computing for providing computational resources and support that have contributed to the research results reported in this publication. 


\end{acknowledgments}


\end{document}